\documentclass{aa}  

\usepackage{graphicx}
\usepackage{txfonts}

\usepackage{natbib}
\bibpunct{(}{)}{;}{a}{}{,} 

\usepackage{multirow}
\usepackage{lscape}
	
\title{North Ecliptic Pole merging galaxy catalogue\thanks{Tables 1 and 4 are only available in electronic form at the CDS via anonymous ftp to cdsarc.u-strasbg.fr (130.79.128.5) or via http://cdsweb.u-strasbg.fr/cgi-bin/qcat?J/A+A/}}

\author{W.~J.~Pearson\inst{\ref{inst:NCBJ}}
		\and L.~E.~Suelves\inst{\ref{inst:NCBJ}}
		\and S.~C.-C.~Ho\inst{\ref{inst:NTHU}}
		\and N.~Oi\inst{\ref{inst:FSD2}}
		\and S.~Brough\inst{\ref{inst:NSW}}
		\and B.~W.~Holwerda\inst{\ref{inst:ULou}}
		\and A.~M.~Hopkins\inst{\ref{inst:AAO}}
		\and T.-C.~Huang\inst{\ref{inst:DSAS}, \ref{inst:ISAS}}
		\and H.~S.~Hwang\inst{\ref{inst:seoul}}
		\and L.~S.~Kelvin\inst{\ref{inst:prince}}
		\and S.~J.~Kim\inst{\ref{inst:NTHU}}
		\and \'{A}.~R.~L\'{o}pez-S\'{a}nchez\inst{\ref{inst:AAO}, \ref{inst:macquarie}, \ref{inst:ASTRO3D}}
		\and K. Ma\l{}ek\inst{\ref{inst:NCBJ}, \ref{inst:LAM}}
		\and C.~Pearson\inst{\ref{inst:OU}, \ref{inst:oxford}, \ref{inst:RAL}}
		\and A.~Poliszczuk\inst{\ref{inst:NCBJ}}
		\and A.~Pollo\inst{\ref{inst:NCBJ}}
		\and V.~Rodriguez-Gomez\inst{\ref{inst:IRA}}
		\and H.~Shim\inst{\ref{inst:DESE}}
		\and Y.~Toba\inst{\ref{inst:kyoto}, \ref{inst:ASIAA}, \ref{inst:RCSCE}}
		\and L.~Wang\inst{\ref{inst:SRON}, \ref{inst:kapteyn}}
}

\institute{National Centre for Nuclear Research, Pasteura 7, 02-093 Warszawa, Poland\label{inst:NCBJ}\\\email{william.pearson@ncbj.gov.pl}
	\and Institute of Astronomy, National Tsing Hua University, 101, Section 2. Kuang-Fu Road, Hsinchu, 30013, Taiwan\label{inst:NTHU}
	\and Faculty of Science Division II, Liberal Arts, Tokyo University of Science,	1-3, Kagurazaka, Shinjuku-ku, Tokyo 162-8601, Japan\label{inst:FSD2}
	\and School of Physics, University of New South Wales, NSW 2052, Australia\label{inst:NSW}
	\and Department of Physics and Astronomy, 102 Natural Science Building, University of Louisville, Louisville KY 40292, USA\label{inst:ULou}
	\and Australian Astronomical Optics, Macquarie University, 105 Delhi Rd, North Ryde, NSW 2113, Australia\label{inst:AAO}
	\and Department of Space and Astronautical Science, Graduate University for Advanced Studies, SOKENDAI, Shonankokusaimura, Hayama, Miura District, Kanagawa 240-0193, Japan\label{inst:DSAS}
	\and Institute of Space and Astronautical Science, Japan Aerospace Exploration Agency, 3-1-1 Yoshinodai, Chuo-ku, Sagamihara, Kanagawa 252-5210, Japan\label{inst:ISAS}
	\and Astronomy Program, Department of Physics and Astronomy, Seoul National University, 1 Gwanak-ro, Gwanak-gu, Seoul 08826, Republic of Korea\label{inst:seoul}
	\and Department of Astrophysical Sciences, Princeton University, 4 Ivy Lane, Princeton, NJ 08544, USA\label{inst:prince}
	\and Department of Physics and Astronomy, Macquarie University, NSW 2109, Australia\label{inst:macquarie}
	\and ARC Centre of Excellence for All Sky Astrophysics in 3 Dimensions (ASTRO-3D)\label{inst:ASTRO3D}
	\and Aix Marseille Univ. CNRS, CNES, LAM, Marseille, France\label{inst:LAM}
	\and The Open University, Milton Keynes, MK7 6AA, UK\label{inst:OU}
	\and RAL Space, Rutherford Appleton Laboratory, Chilton, Didcot, Oxfordshire OX11 0QX, UK\label{inst:oxford}
	\and Oxford Astrophysics, University of Oxford, Keble Rd, Oxford OX1 3RH, UK\label{inst:RAL}
	\and Instituto de Radioastronom\'{i}a y Astrof\'{i}sica, Universidad Nacional Aut\'{o}noma de M\'{e}xico, A.P. 72-3, 58089 Morelia, Mexico\label{inst:IRA}
	\and Department of Earth Science Education, Kyungpook National University, Daegu 41566, Republic of Korea\label{inst:DESE}
	\and Department of Astronomy, Kyoto University, Kitashirakawa-Oiwake-cho, Sakyo-ku, Kyoto 606-8502, Japan\label{inst:kyoto}
	\and Academia Sinica Institute of Astronomy and Astrophysics, 11F of Astronomy-Mathematics Building, AS/NTU, No.1, Section 4, Roosevelt Road, Taipei 10617, Taiwan\label{inst:ASIAA}
	\and Research Center for Space and Cosmic Evolution, Ehime University, 2-5 Bunkyo-cho, Matsuyama, Ehime 790-8577, Japan\label{inst:RCSCE}
	\and SRON Netherlands Institute for Space Research, Landleven 12, 9747 AD, Groningen, The Netherlands\label{inst:SRON}
	\and Kapteyn Astronomical Institute, University of Groningen, Postbus 800, 9700 AV Groningen, the Netherlands\label{inst:kapteyn}
}

\date{Received 07 April 2021 / Accepted 20 February 2022}

\abstract{}
{We aim to generate a catalogue of merging galaxies within the 5.4 sq. deg. North Ecliptic Pole over the redshift range $0.0 < z < 0.3$. To do this, imaging data from the Hyper Suprime-Cam are used along with morphological parameters derived from these same data.}
{The catalogue was generated using a hybrid approach. Two neural networks were trained to perform binary merger non-merger classifications: one for galaxies with $z < 0.15$ and another for $0.15 \leq z < 0.30$. Each network used the image and morphological parameters of a galaxy as input. The galaxies that were identified as merger candidates by the network were then visually checked by experts. The resulting mergers will be used to calculate the merger fraction as a function of redshift and compared with literature results.}
{We found that 86.3\% of galaxy mergers at $z < 0.15$ and 79.0\% of mergers at $0.15 \leq z < 0.30$ are expected to be correctly identified by the networks. Of the 34\,264 galaxies classified by the neural networks, 10\,195 were found to be merger candidates. Of these, 2109 were visually identified to be merging galaxies. We find that the merger fraction increases with redshift, consistent with literature results from observations and simulations, and that there is a mild star-formation rate enhancement in the merger population of a factor of $1.102 \pm 0.084$.}
{}

\keywords{Catalogs -- Galaxies: interactions -- Galaxies: evolution -- Methods: data analysis -- Galaxies: statistics}

\begin{document}
\maketitle

\section{Introduction}\label{sec:intro}
Galaxy mergers underpin our current understanding of how galaxies grow and evolve. In the current cold dark matter paradigm, dark matter halos assemble hierarchically. This results in the baryonic constituents of the dark matter halos also merging. The result is a larger galaxy living in the heart of a larger dark matter halo \citep[e.g.][]{2014ARA&A..52..291C, 2015ARA&A..53...51S}.

Numerous studies have looked at how galaxy-galaxy mergers influence the star-formation rate (SFR) or active galactic nuclei (AGN) activity of both the progenitor and descendant galaxies. The merger and SFR connection was raised when early infrared observations found that the majority of infrared-bright galaxies were merging. The link between infrared-bright galaxies and high SFRs resulted in the conclusion that galaxy mergers can trigger periods of highly enhanced SFRs and starbursts \citep[e.g.][]{1985MNRAS.214...87J, 1996ARA&A..34..749S, 2012MNRAS.421.1539N}. The increase in SFRs during a merger event has been seen in more recent works, although not all galaxy mergers are seen with highly enhanced SFRs that would be considered starbursts.
	
The constituent galaxies of a merger are found to play a role in the strength of star-formation enhancement. Interactions between two spiral galaxies have been shown to have an enhanced SFR, when compared to non-mergers, while little enhancement is seen when at least one of the merging galaxies is elliptical \citep{2011A&A...535A..60H}. The strength of the interaction also influences star-formation, with galaxies whose projected distance to the nearest neighbour is less than a tenth of the virial radius of the nearest neighbour experiencing greater increases in specific SFRs, up to a factor of 4 \citep{2011A&A...535A..60H}. Post-merger galaxies are also seen to have their SFR increase by a factor of approximately 4 when compared to a non-merging control sample \citep{2013MNRAS.435.3627E}. The mass of the interacting galaxies is also likely to contribute to the star-formation enhancement, with galaxies with stellar masses below 10$^{11}$~M$_{\odot}$ showing a greater enhancement than more massive mergers. Indeed, major mergers of dwarf galaxies are found to have similar star-formation enhancement to more massive major mergers \citep{2015ApJ...805....2S}.
	
Other studies have found weaker enhancement in star-formation during a merger. In \citet{2015MNRAS.454.1742K}, galaxy mergers are found to typically show mild star-formation enhancement, a factor of approximately 1.9 at most, with many merging systems showing no enhancement. These enhancements, or lack thereof, were determined by dividing the SFR of a merging system with the median SFR of that system's control group. Similarly, \citet{2019A&A...631A..51P} also found mild enhancement, with the average SFR in mergers to be only a factor of 1.2 higher than the average SFR in non-mergers. However, the \citet{2019A&A...631A..51P} merger sample is likely to be highly contaminated by non-mergers due to their selection from only a neural network. Reductions in SFRs during galaxy mergers can be seen in low mass (stellar mass $< 10^{9}$~M$_{\odot}$) secondary galaxies of minor mergers \citep{2015MNRAS.452..616D, 2016MNRAS.455.4013D}. Many dwarf starbursts, such as blue compact dwarf galaxies, appear to be a consequence of the strong interactions or mergers of even smaller entities. However, these features are only observed when deep images and complementary spectroscopic and/or radio data are available \citep{2010A&A...521A..63L, 2012ApJ...748L..24M, 2020ApJ...900..152Z}. What merger studies do agree on, however, is that not all galaxy mergers are undergoing a starburst at the time of observation but starbursts are more common in mergers than non-mergers \citep{2008AJ....135.1877E, 2011A&A...535A..60H, 2012MNRAS.426..549S, 2013MNRAS.435.3627E, 2013MNRAS.433L..59P, 2015MNRAS.454.1742K, 2015ApJ...805....2S, 2019A&A...631A..51P}.

These observational findings agree with what is seen in simulations. Zoom-in simulations of merging galaxies allow the SFR to be closely tracked during an entire simulated merger with fine time-resolution. Such simulations indicate that galaxies go through short periods of highly enhanced star-formation \citep[e.g.][]{2006MNRAS.373.1013C, 2011ApJ...730....4B, 2013MNRAS.430.1901H, 2015A&A...575A..56B, 2016MNRAS.462.2418S, 2019MNRAS.485.1320M, 2019MNRAS.490.2139R}. These are typically seen around first close passage and coalescence of the merging galaxies. Thus, only short periods of a galaxy merger are able to be observed to have highly enhanced SFRs resulting in real galaxies typically being observed while only experiencing mild SFR enhancement.

Integral field observations have allowed resolved star-formation, rather than global star-formation, to be traced in mergers. With such observations, the merger triggered star-formation has been seen to primarily occur in the centre of a galaxy while the outer regions of the interacting galaxies show enhancement or suppression \citep{2019MNRAS.482L..55T}, with the enhancement or suppression being dependent on the merger period \citep{2019ApJ...881..119P}.

High infrared emission can also be linked with AGN activity, where the AGN are known to heat the dust that surrounds them, emitting strongly in the infrared. Galaxy mergers have been seen to drive material onto a central black hole of a galaxy, feeding the AGN and resulting in increased activity \citep[e.g.][]{1985AJ.....90..708K, 2011ApJ...743....2S, 2012A&A...538A..15H, 2014AJ....148..137L, 2014MNRAS.441.1297S, 2014MNRAS.437.2137S, 2017MNRAS.464.3882W, 2018PASJ...70S..37G, 2019MNRAS.487.2491E, 2020A&A...637A..94G}. However, this interpretation is contested, with a number of studies finding similar fractions of AGN in and out of galaxy mergers \citep[e.g.][]{2012ApJ...744..148K, 2016ApJ...830..156M, 2021arXiv210105000S}. This contention may be due to differences in the type of selected AGN \citep[e.g. obscured or unobscured;][]{2010ApJ...716L.125K, 2015ApJ...814..104K} or differing merger identification methods \citep{2021arXiv210615618L}.

The merger rate and fraction in the Universe is not constant with redshift. Both observations and simulations typically agree that the fraction and rate of galaxy mergers was higher in the earlier Universe and has decreased as the Universe has aged \citep[e.g.][]{2002ApJ...565..208P, 2004ApJ...617L...9L, 2007ApJS..172..320K, 2009A&A...498..379D, 2011ApJ...742..103L, 2013MNRAS.431.2661C, 2013A&A...553A..78L, 2014MNRAS.445.1157C, 2015MNRAS.449...49R, 2017MNRAS.470.3507M, 2017MNRAS.464.1659Q, 2018MNRAS.477.1822M, 2019ApJ...876..110D, 2019A&A...631A..51P, 2020ApJ...895..115F, 2021MNRAS.501.3215O}. The observationally determined merger fraction and rate evolutions use different selection methods, providing a firm determination of the increase of these two values with redshift. However, the exact evolution of the merger fraction and merger rate differ between different studies. Indeed, the simulations also do not agree on the evolution of these two quantities. The Horizon-AGN cosmological simulation \citep{2014MNRAS.444.1453D} finds no evolution of the merger fraction with redshift \citep{2015MNRAS.452.2845K}, unlike other simulations that find an increase with redshift \citep{2015MNRAS.449...49R, 2017MNRAS.464.1659Q}. There is also observational evidence that the merger fraction may reduce above $z \approx 2$ for intermediate mass galaxies \citep[M$_{\star}$ between 10$^{9}$ and 10$^{10}$~M$_{\odot}$][]{2008MNRAS.386..909C}. The difference in the evolution of the merger rate when compared to the evolution of the merger fraction may be resolved by using an evolving merger timescale instead of a fixed merger timescale \citep{2017MNRAS.468..207S}.

Merging galaxies are traditionally selected by identifying close pairs, that is finding galaxies that are close both on the sky and in redshift \citep[e.g.][]{2000ApJ...530..660B, 2005AJ....130.1516D, 2014MNRAS.444.3986R, 2018MNRAS.475.5133R, 2019ApJ...876..110D}, or morphologically disturbed systems, identified either visually or through parametric and non-parametric statistics \citep[e.g.][Kim et al. in Prep]{2000AJ....119.2645B, 2000ApJ...529..886C, 2003AJ....126.1183C, 2008MNRAS.389.1179L}. For the latter technique, the majority of mergers are only identifiable for part of the merger time \citep{2010MNRAS.404..590L, 2010MNRAS.404..575L} and any study of merger rates or fractions with such identifications assumes that the scatter into and out of the selection is approximately equal. Visual selection, in particular, is a time intensive task which limits the sample size of merging galaxies that can be identified while the classifications can be difficult to reproduce and can be incomplete \citep{2015ApJS..221....8H}. This visual selection is also biased towards mergers that are closer to a pericentric passage where the morphological disturbance caused by the interaction is more visible \citep[e.g.][]{2020MNRAS.492.2075B}. More recent developments have allowed the detection of merging galaxies using machine learning which is orders of magnitude faster than visual selection \citep[e.g.][]{2018MNRAS.479..415A, 2019MNRAS.490.5390B, 2019ApJ...872...76N, 2019A&A...631A..51P, 2019A&A...626A..49P, 2019MNRAS.483.2968W, 2020ApJ...895..115F, 2020A&A...644A..87W}. However, such identifications are known to suffer from impurity of the merger sample \citep[e.g. as shown by][]{2021MNRAS.504..372B} and are limited by the quality of the training sample. Here we aim to obtain a clean sample of merging systems which will allow detailed follow-up studies of a statistically large number of galaxy mergers. Thus, we combine the speed of machine learning identification with a accuracy of visual classification.

This paper presents a large catalogue of merging galaxies with redshifts between 0.0 and 0.3 that is ideally suited for studying the link between merging galaxies and rarer astrophysical phenomena, such as AGN. The presented catalogue is for galaxies within the North Ecliptic Pole (NEP), a 5.4 sq. deg. area that has been well studied in numerous wavelength ranges \citep{2021MNRAS.500.4078K}, including infrared data from \textit{AKARI} \citep{2007PASJ...59S.369M, 2012A&A...548A..29K}, optical data from the Hyper Suprime-Cam \citep[HSC;][]{2017PKAS...32..225G, 2018PASJ...70S...3F, 2018PASJ...70...66K, 2018PASJ...70S...2K, 2018PASJ...70S...1M, 2021MNRAS.500.5024O} and X-ray data from \textit{Chandra} \citep{2015MNRAS.446..911K}. This allows for studies correlating galaxy mergers with rare phenomena to be undertaken. The NEP will also be used as the location for a deep \textit{Euclid} field \citep{2011arXiv1110.3193L}. Thus the objects within the catalogue will have high quality near-infrared images taken in the near future. This catalogue will provide an excellent training sample for automated detection of further mergers throughout the \textit{Euclid} coverage.

The catalogue presented in this work was generated using a hybrid deep learning - human approach, as proposed by \citet{2021MNRAS.504..372B}. Deep learning techniques, applied to imaging and morphological data, were used to generate a sample of merger candidates. These merger candidates were then visually inspected by professional astronomers to create a final catalogue of galaxy mergers. The paper is structured as follows. Section \ref{sec:data} describes the data used to generate this catalogue. Section~\ref{sec:dl} discusses deep learning and the neural networks used to generate the merger candidates along with the human verification process. Section~\ref{sec:results} presents the results of the merger identification and Sect.~\ref{sec:discuss} presents discussion on these classifications. We summarise our work in Sect. \ref{sec:conc}.

\section{Data}\label{sec:data}
\subsection{Imaging data}
For the training data, we used imaging data from the HSC Subaru Strategic Program (HSC-SSP) Data Release 2 \citep[DR2;][]{2018PASJ...70S...4A, 2019PASJ...71..114A}. The galaxies used for training were selected using $r$-band data (see Sect. \ref{sec:known}) and so HSC-SSP wide field $r$-band imaging was used. Within the HSC-SSP, the wide field $r$-band magnitude 5$\sigma$ limit is 26.2 AB mag. The morphological parameters were also derived from the $r$-band HSC-SSP data using \texttt{statmorph} \citep{2019MNRAS.483.4140R}.

For identifying galaxy mergers within NEP, HSC data from the HSC survey of NEP were used \citep[HSC-NEP;][]{2017PKAS...32..225G, 2021MNRAS.500.5024O}. Here we again used the $r$-band data, which reaches a median 5$\sigma$ depth of 27.3 AB mag, to match the band used for the training data. This choice of band is despite the HSC-NEP $r$-band having poorer seeing than other HSC-NEP optical bands: 1.26~arcsec in the $r$-band compared to 0.68~arcsec in the $g$-band \citep{2021MNRAS.500.5024O}. Galaxy positions and magnitudes were derived by \citet{2021MNRAS.500.5024O} using the HSC data analysis pipeline version 4.0.1 \citep{2018PASJ...70S...5B}. The photometric redshifts for the NEP galaxies were derived in \citet{2021MNRAS.502..140H} using the Canada France Hawaii Telescope MegaPrime u-band \citep{2003SPIE.4841...72B, 2014A&A...566A..60O, 2020MNRAS.498..609H}, HSC $g$, $r$, $i$, $z$, and $y$-bands, and the Spitzer Infrared Array Camera bands 1 and 2 \citep{2004ApJS..154...10F, 2018ApJS..234...38N} using \texttt{LePhare} \citep{1999MNRAS.310..540A, 2006A&A...457..841I}. The photometric redshifts have a weighted dispersion of $\sigma_{\Delta z/(1+z)} = 0.053$ and catastrophic error fraction of 11.3\%. Spectroscopic redshifts were derived from optical spectroscopy \citep{2013ApJS..207...37S, 2017PASJ...69...70O, 2018cwla.conf..371K, 2018A&A...618A.101O}. The galaxy sample that was checked for mergers were chosen where their photometric redshift, or spectroscopic redshift where available, is less than $z = 0.30$. Above this redshift, the quality of the neural networks used to identify the galaxy mergers rapidly deteriorated. Of the 34\,264 galaxies from HSC-NEP with $z < 0.30$, 736 have spectroscopic redshifts and the remaining 33\,528 have photometric redshifts. Morphological parameters were again derived using \texttt{statmorph} using the $r$-band images and segmentation maps were created using \texttt{SExtractor} \citep{1996A&AS..117..393B}.

\subsection{Morphological parameters}
To supplement the imaging data, morphological parameters of the galaxies were also used to help identify galaxy mergers. The morphological parameters used in this work were all derived from the HSC $r$-band images using the \texttt{statmorph} python package. 
These parameters are described below.

The concentration \citep[C;][]{1985ApJS...59..115K, 1994ApJ...432...75A, 2000AJ....119.2645B, 2003ApJS..147....1C} describes the ratio between amount of light towards the centre of a galaxy with the amount of light within a larger radius. The \texttt{statmorph} package follows \citet{2004AJ....128..163L} and compares the ratio of the radius that contains 20\% of the light and the radius that contains 80\% of the light. Larger values of C indicate that more light is concentrated in the centre of the galaxy.

The asymmetry \cite[A;][]{1996ApJS..107....1A, 2000ApJ...529..886C} measures the rotational symmetry of a galaxy, the calculation of which again follows \citet{2004AJ....128..163L}. An image is rotated by 180$^{\circ}$ and this rotated image is subtracted from the original image. The residual values in the pixels are summed to give the final value of asymmetry. Larger values of asymmetry indicate that a galaxy is less rotationally symmetric.

The smoothness \citep[S;][]{1999ApJS..122..109T, 2003ApJS..147....1C} determination in \texttt{statmorph} follows the definition of \citet{2004AJ....128..163L}. A smoothed image is created by applying a smoothing filter of fixed size to the original image. The new image is subtracted from the original image, leaving only the high frequency disturbances. This residual image is then summed, with higher values indicating a less smooth (more clumpy) galaxy.

The Gini coefficient \citep{2003ApJ...588..218A} describes the distribution of light among pixels. If the Gini value is 1, all the light is in a single pixel, while if Gini is 0, all the light is shared equally across all pixels. Gini provides a description of how concentrated the light is within an image, independent of the spatial distribution of that light. Gini is calculated following \citet{2004AJ....128..163L} by determining the mean of the absolute difference between all pixels.

M$_{20}$ \citep{2004AJ....128..163L} describes the second-order moment of the brightest 20\% of a galaxy's pixels normalised by the second-order moment of the entire galaxy. Again, \texttt{statmorph} follows \citet{2004AJ....128..163L} and calculates the second-order moment by summing the distance of a pixel to the centre of a galaxy multiplied by the flux of the pixel. Less negative M$_{20}$ implies a galaxy is more concentrated, although there is no requirement that this concentration is in the centre of a galaxy.

The Gini-M$_{20}$ bulge parameter \citep[GMB;][]{2015MNRAS.454.1886S, 2019MNRAS.483.4140R} is five times the perpendicular distance from a galaxy to the line that separates early and late type galaxies in the Gini-M$_{20}$ plane. The definition used by \texttt{statmorph} is that of \citet{2019MNRAS.483.4140R}:
\begin{equation}\label{eqn:gmb}
	GMB = -0.693~M_{20} + 4.95~Gini - 3.96.
\end{equation}
Larger GMB imply a greater bulge domination while a lower GMB implies greater disk domination. GMB is less sensitive to dust and mergers than M$_{20}$, concentration or the S\'{e}rsic index \citep{2015MNRAS.454.1886S}.

Gini-M$_{20}$ merger parameter \citep[GMM;][]{2004AJ....128..163L,2008ApJ...672..177L,  2015MNRAS.454.1886S, 2019MNRAS.483.4140R} is similar to GMB. It is the position along a line that lies perpendicular to the line that separates merging from non-merging galaxies in the Gini-M$_{20}$ plane. Thus, GMM is defined as:
\begin{equation}\label{eqn:gmm}
	GMM = 0.139~M_{20} + 0.990~Gini - 0.327.
\end{equation}
This formulation adopts the Gini-M$_{20}$ merger classification of \citet{2008ApJ...672..177L}, which should allow better application over a larger range of redshifts than the \citet{2004AJ....128..163L} classification \citep{2015MNRAS.451.4290S, 2015MNRAS.454.1886S}.

The multimode statistic (M) is the ratio of the area between the two brightest regions of a galaxy \citep{2013MNRAS.434..282F, 2016MNRAS.458..963P}. The bright regions are determined by cutting at a flux threshold and finding the two brightest regions above the threshold. This is repeated with different flux thresholds and the multimode statistic is then the largest ratio. If this ratio is closer to 1, the object is more likely to contain two nuclei. 

The intensity statistic (I) is similar to the multimode. Here, the ratio of the fluxes of the brightest two regions is taken \citep{2013MNRAS.434..282F, 2016MNRAS.458..963P}. The two brightest regions are defined by finding local maxima of a smoothed image of the galaxy, identified by following the gradient of the flux. If the intensity is closer to 1, the galaxy is more clumpy.

The deviation statistic (D) is calculated by determining the distance between the galaxy intensity centroid and the centre of the brightest region \citep{2013MNRAS.434..282F, 2016MNRAS.458..963P}. A high value for deviation implies that the galaxy is clumpy and the bright regions are significantly separated from the intensity centroid.

The ellipticity asymmetry (Eli A) and centroid (Eli Cen) are the ellipticity of the galaxy relative to the point that minimises the asymmetry or relative to the centroid. Similarly, the elongation asymmetry (Elo A) and centroid (Elo Cen) are the elongation of the source relative to the point that minimises asymmetry or relative to the centroid of the galaxy \citep{2019MNRAS.483.4140R}.

The S\'{e}rsic index ($n$) is the best fit power law index for the S\'{e}rsic profile \citep{1963BAAA....6...41S, 2005PASA...22..118G} that has been fitted to the light profile of an entire galaxy. Larger S\'{e}rsic indices imply a more bulge dominated galaxy, although it is possible to find bulge dominated galaxies with low S\'{e}rsic indices \citep{2003AJ....125.2936G}. 
The S\'{e}rsic amplitude (SA) is the amplitude of the S\'{e}rsic profile at the effective (half-light) radius while the S\'{e}rsic ellipticity (SE) is the ellipticity of the profile.

\begin{figure*}
	\resizebox{0.95\hsize}{!}{\includegraphics{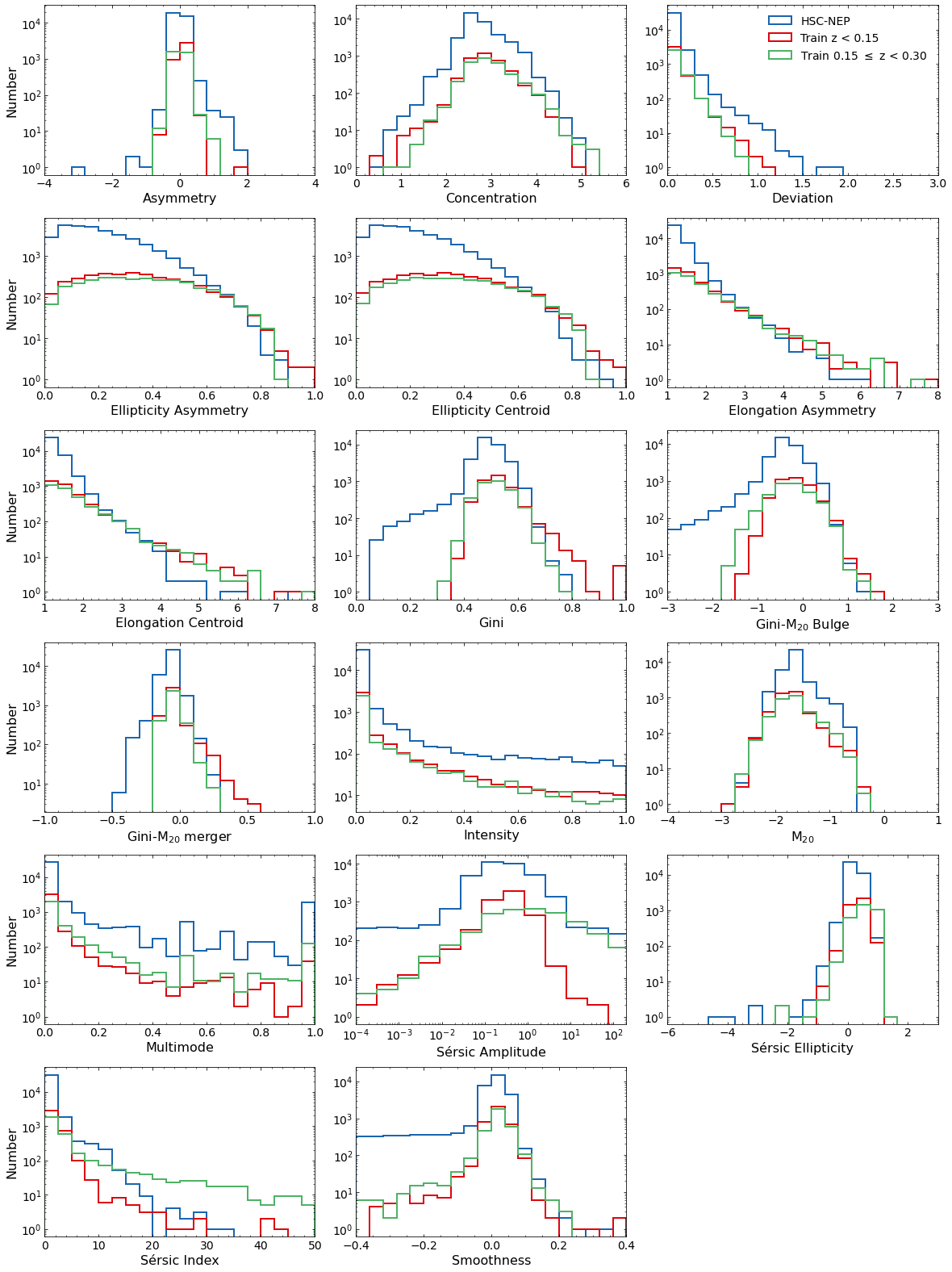}}
	\caption{Distributions of morphological parameters for NEP-SCP galaxies (blue), training data for the $z < 0.15$ network (red) and training data for the $0.15 \leq z < 0.30$ network (green). The range shown is that used for training (see Sect. \ref{subsec:NNtvt} and Table \ref{tab:morph})}
	\label{fig:morph}
\end{figure*}

While the above parameters are not all completely independent of one another, for example the S\'{e}rsic index will be monotonically related to the concentration if a S\'{e}rsic profile is a good description of a galaxy's light profile \citep{2001AJ....122.1707G, 2020ApJ...903...97S}, they do all individually describe slightly different properties of a galaxy. However, GMB and GMM are both derived from combinations of Gini and M$_{20}$ and so will not be independent of combinations of Gini and M$_{20}$. Thus a neural network may be able to discern differences between these parameters that are subtle but aid in merger identification. The morphological parameters for the HSC-NEP galaxies are presented in Table \ref{tab:NEP-morph} and Fig. \ref{fig:morph}.

\begin{table*}
	\caption[]{Ten rows of the morphology catalogue for galaxies in NEP}
	\label{tab:NEP-morph}
	\centering
	\begin{tabular}{cccccccccc}
		\hline
		HSC\_ID & A & C & D & Eli A & Eli Cen & Elo A & Elo Cen & Gini & GMB \\
		\hline
		79666794322744899 & 0.033 & 2.735 & 0.046 & 0.194 & 0.194 & 1.241 & 1.24 & 0.532 & -0.108 \\
		79671166599467769 & 0.036 & 2.988 & 0.012 & 0.291 & 0.291 & 1.411 & 1.411 & 0.519 & -0.130 \\
		79671179484351321 & -0.035 & 2.597 & 0.045 & 0.094 & 0.094 & 1.104 & 1.103 & 0.513 & -0.231 \\
		79218331017565336 & 0.017 & 3.243 & 0.027 & 0.208 & 0.208 & 1.263 & 1.263 & 0.553 & 0.103 \\
		80093924525370378 & -0.271 & 2.574 & 0.030 & 0.065 & 0.062 & 1.007 & 1.066 & 0.416 & -1.175 \\
		79671029160501625 & -0.091 & 2.683 & 0.061 & 0.439 & 0.44 & 1.783 & 1.786 & 0.508 & -0.397 \\
		80093108481580765 & 0.023 & 2.500 & 0.038 & 0.061 & 0.061 & 1.065 & 1.065 & 0.474 & -0.468 \\
		79670625433569331 & 0.023 & 2.662 & 0.023 & 0.285 & 0.285 & 1.398 & 1.398 & 0.474 & -0.414 \\
		80093112776555761 & 0.021 & 2.843 & 0.037 & 0.094 & 0.094 & 1.104 & 1.104 & 0.525 & -0.129 \\
		79666506559929228 & -0.03 & 2.372 & 0.008 & 0.256 & 0.255 & 1.343 & 1.343 & 0.448 & -0.624 \\
		... & ... & ... & ... & ... & ... & ... & ... & ... & ... \\
		\hline
	\end{tabular}
	\begin{tabular}{ccccccccc}
		\hline
		HSC\_ID & GMM & I & M$_{20}$ & M & SA & SE & $n$ & S \\
		\hline
		79666794322744899 & -0.046 & -1.761 & 0.02 & 0.507 & 0.159 & 1.242 & 0.007 & 0.033 \\
		79671166599467769 & -0.065 & -1.816 & 0.004 & 0.706 & 0.385 & 1.702 & 0.024 & 0.036 \\
		79671179484351321 & -0.058 & -1.716 & 0.008 & 0.283 & 0.138 & 1.254 & 0.032 & -0.035 \\
		79218331017565336 & -0.046 & -1.916 & 0.001 & 4.547 & 0.229 & 2.05 & 0.014 & 0.017 \\
		80093924525370378 & -0.060 & -1.045 & 1.000 & 0.098 & 0.065 & 1.000 & -0.960 & -0.271 \\
		79671029160501625 & -0.034 & -1.511 & 1.000 & 0.236 & 0.579 & 0.744 & 0.003 & -0.091 \\
		80093108481580765 & -0.088 & -1.656 & 0.000 & 0.954 & 0.126 & 0.927 & 0.053 & 0.023 \\
		79670625433569331 & -0.099 & -1.735 & 0.004 & 1.892 & 0.362 & 1.134 & 0.005 & 0.023 \\
		80093112776555761 & -0.054 & -1.777 & 0.005 & 0.858 & 0.188 & 1.535 & 0.005 & 0.021 \\
		79666506559929228 & -0.109 & -1.617 & 0.005 & 0.343 & 0.333 & 0.711 & -0.001 & -0.030 \\
		... & ... & ... & ... & ... & ... & ... & ... & ... \\
		\hline
	\end{tabular}
\end{table*}

\subsection{Known mergers and non-mergers}\label{sec:known}
For supervised learning, it is necessary to have a sample of objects with known labels, here merger or non-merger, to use to train a machine learning algorithm. For this we used the same sample of merging and non-merging galaxies used as a training set in \citet{2019A&A...631A..51P}. This training sample was selected in \citet{2019A&A...631A..51P} using results from the GAMA-KiDS Galaxy Zoo project \citep[][Kelvin et al. in prep]{2008MNRAS.389.1179L, 2009A&G....50e..12D, 2013Msngr.154...44D, 2013ExA....35...25D, 2019AJ....158..103H} along with an A-S cut \citep{2003ApJS..147....1C} with the A and S parameters used in this selection derived from KiDS $r$-band imaging. These galaxies have a redshift below 0.15. \citet{2019A&A...631A..51P} define a merger to be a galaxy with \texttt{mergers\_neither\_frac} from Galaxy Zoo to be less than 0.5, that is less than half the citizen scientists determined a galaxy had no evidence of tidal tails or evidence of a merger, and had $A > 0.35S + 0.02$. Non-mergers were defined by \citet{2019A&A...631A..51P} to have \texttt{mergers\_neither\_frac} $>$ 0.5 and $A < 0.35S + 0.02$.

\begin{figure}
	\resizebox{\hsize}{!}{\includegraphics{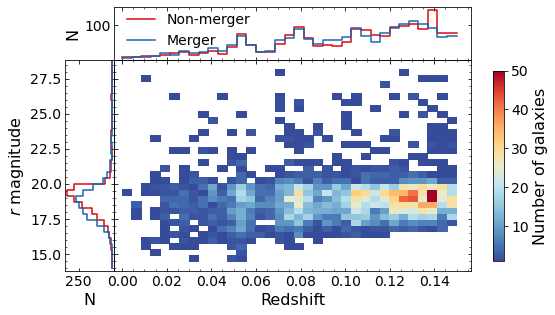}}
	\caption{Density plot of HSC-SSP $r$ band magnitude against GAMA spectroscopic redshift for the 3366 training galaxies, both merging and non-merging, binned by $r$ band magnitude and redshift. Blue bins have fewer galaxies and red bins have a larger number of galaxies. Left and upper panels show the $r$ magnitude and redshift distributions, respectively, of the non-merger (red) and merger (blue) galaxies.}
	\label{fig:rmag}
\end{figure}

We limited the sample of galaxies we used to those that lie in both the GAMA-KiDS coverage as well as the HSC-SSP coverage so that all training objects have HSC data available. As such, the sample is smaller than the sample used by \citet{2019A&A...631A..51P} as the HSC-SSP DR2 does not cover all of the area covered by GAMA-KiDS. The resulting sample, which is intentionally class balanced, is 1\,683 merging galaxies with 1\,683 non-merging galaxies. This balance was achieved by randomly removing galaxies from the larger class until there were the same number of merging and non-merging galaxies. The HSC $r$ magnitude distribution for the whole training sample is presented as a function of redshift in Fig. \ref{fig:rmag}. For use while training the networks that will be employed in this work, $r$-band cutouts of 128$\times$128 pixels, corresponding to approximately 21.5$\times$21.5 arcsec, were made. The morphological parameters used within the networks were derived from these cutouts using the \texttt{statmorph} Python package. The square root of the HSC variance maps were used as the weight maps for \texttt{statmorph}. The morphological parameters can be seen in Fig. \ref{fig:morph}.

As the selection of merging galaxies was aided by the A-S cut, it is likely that the non-merging galaxies have little or no visible structure, a result of the A-S cut splitting featured and non-featured galaxies \citep{2003ApJS..147....1C}. However, as the merging galaxies are visually selected with Galaxy Zoo, these are likely to be galaxies with the visual appearance of mergers. As a result, the non-mergers selected by a network trained with this data have the potential to be selected due to their lack of features. This provides further justification of visual confirmation of the mergers selected by the neural networks used in this work, beyond the non-merger contamination expected from any machine learning technique.

This work also identified galaxies at redshifts between 0.15 and 0.30. For this, the galaxies used at $z < 0.15$ were augmented to appear like higher redshift galaxies. This was done as there is not a sample of known galaxy mergers between these redshifts in GAMA-KiDS or NEP. A random redshift between 0.15 and 0.30 was selected and assigned to each galaxy and the apparent $r$-band magnitude of each galaxy dimmed to match that redshift. For any galaxies whose new apparent magnitude was greater than 26 AB, the approximate $r$-band magnitude limit for the HSC-SSP, 0.15 was added to the original redshift of the galaxy and the apparent magnitude re-calculated. Galaxies whose apparent magnitudes were still above 26 AB were removed. The physical resolutions of the remaining galaxies were adjusted to match their new redshift. Galaxy cutouts that were 256$\times$256 pixels were rebinned to reduce blank space around the resized, 128$\times$128 pixel images that were used for training the networks. Synthetic, Gaussian noise was then added to the image, which also filled any blank space around the resized images. The standard deviation of the synthetic noise was determined by calculating the standard deviation of the original image, before redshift dimming, after 3$\sigma$ clipping 100 times. The clipping derived noise is approximately a factor of 10 larger than the HSC weight maps (that is the maps of the 1$\sigma$ values of each pixel). This larger noise will not be a perfect representation of the real images and so provides further requirement for a visual check to confirm the merger candidates from the neural networks are real mergers. The size of the images was still 128$\times$128 pixels and the synthetic noise was used to fill the empty space around the re-binned image. Segmentation maps were generated using \texttt{SExtractor} and morphological parameters re-derived using \texttt{statmorph}, using the square root of appropriately scaled version of the HSC variance maps as the weight maps, and can be seen in Fig. \ref{fig:morph}. The scaling of the weight maps includes both the resolution scaling and synthetic noise contribution. The higher noise may also influence the morphological parameters from \texttt{statmorph}. This sample was again class balanced by random removal of galaxies in the larger class.

\begin{figure}
	\resizebox{\hsize}{!}{\includegraphics{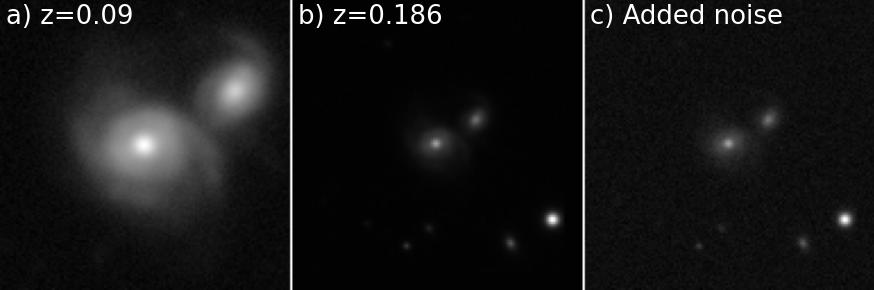}}
	\caption{Example of augmentation of a galaxy to a higher redshift. The original galaxy is shown in panel (a), the physically rebinned and dimmed galaxy is shown in panel (b) and the final image, with added Gaussian noise, is shown in panel (c). All panels have asinh scaling and are 128$\times$128 pixels.}
	\label{fig:ex_aug}
\end{figure}

K-correction was not applied to these redshifted images. Using the average spectral energy distribution template of \citet{2001ApJ...556..562C}, an increase in redshift by 0.15, from $z = 0.075$ to $z = 0.225$, would require a K-correction of approximately a factor of 1 for the $r$-band (i.e. no correction is required). The same factor is seen using the \citet{2008ApJ...682..985W} template while the average SWIRE Template Library template \citep{2007ApJ...663...81P} has a factor of $\sim$0.9. The exact K-correction will differ between specific galaxies but this difference is not expected to be large. For the same redshift change, the magnitude is changed by approximately 2.5 or the flux is changed by a factor of approximately 10.

For generating synthetic seeing in the redshifted galaxies, four options were considered. The original image could be deconvolved with the point-spread function (PSF), the image resized and this new image re-convolved with the PSF, which is not possible as deconvolving a noisy image results in the destruction of the image. A second PSF could be calculated to re-create the original PSF in the rescaled images. The resized image could be convolved with the PSF, which would result in over-distortion. Or no alteration could be done, which would result in an under-distorted image. Here we performed no convolution and accept that the resulting images will be under-distorted. The PSF of the original, non-redshifted image was used within \texttt{statmorph} when deriving the morphological parameters. As this is likely to introduce errors in the morphological parameters and cause the images to not be a perfect representation of the real images, this provides further requirement for visual confirmation of the merger candidates from the neural network.

\subsection{Mass completeness}\label{subsec:mass}
For application of the merger sample derived in this work (Sect. \ref{sec:discuss}) it is necessary to determine the mass completeness limit. This mass completeness estimate was done empirically following \citet{2010A&A...523A..13P}:
\begin{equation}
log(M_{lim}) = log(M_{\star}) - 0.4(r_{lim} - r),
\end{equation}
where M$_{\star}$ is the stellar mass of a galaxy in M$_{\odot}$, $r_{lim}$ is the limiting $r$-band magnitude, here set to 26, $r$ is the measured $r$-band magnitude of the galaxy and M$_{lim}$ is the lowest mass that can be observed for this object at the $r$-band magnitude limit. The limiting mass within a redshift bin is then the M$_{lim}$ value that 90\% of the faintest 20\% of galaxies have masses below. The masses of each galaxy were determined at the same time as their photometric redshifts through spectral energy distribution fitting using \texttt{LePhare} \citep{1999MNRAS.310..540A, 2006A&A...457..841I, 2021MNRAS.502..140H}. While this calculation of the completeness limit was for the $I$-band in \citet{2010A&A...523A..13P}, we find that using the $r$-band provides a more conservative mass limit. As the galaxy selection, morphologies, and classifications are based on $r$-band data, it was decided to use the more conservative $r$-band mass limit over the $I$-band limit.

\section{Deep learning}\label{sec:dl}
Deep learning is a subset of machine learning that aims to loosely mimic how biological neural networks process data. This work employs a convolutional neural network (CNN) combined with a traditional neural network. CNNs are designed to better process multi-dimensional data, such as images, by reducing the number of trainable parameters within a network. Here, we specifically perform supervised learning, where the truth values for the training data are known. The training data are typically sub-divided into three subsets: a `training set', which typically contains 70\% to 90\% of the training data, used to train the network; a `validation set', which typically contains 5\% to 15\% of the training data, used to evaluate the performance of a network as it is trained; and a `test set', which again typically contains 5\% to 15\% of the training data, that are not shown to a network during training and only used once to test a network once training is complete. The exact split between the three data sets is a matter of choice and varies between studies: here we use 80\% for the training set, 10\% for the validation set and 10\% for the test set. For ease of communication, neural networks that are not CNNs will be referred to as fully connected networks (FCN).

\subsection{Neural network architecture}\label{subsec:NNa}
For this work, we emploied a hybrid neural network containing a FCN and a CNN, the output of which are combined to form a final result \citep[e.g.][]{ZhouIEEE2017, 2019A&A...624A.102D}. The FCN side of the network has morphological parameters passed into it while the CNN has an $r$-band image of the galaxy being classified passed into it. Each part of the network could be used to determine if a galaxy is a merger or non-merger, however we found that the combination of both provides better results (see Sect. \ref{subsec:nn}). Unless otherwise stated, the hyper-parameters for the layers, activations, batch normalisations, drop out and optimiser were left at the TensorFlow default values.

The FCN side comprises two layers containing 128 neurons. The output layer of this network comprises two neurons, one each for the merger and non-merger probabilities. Rectified linear units \citep[ReLU;][]{nair2010rectified} are used for activation in the two layers of 128 neurons while softmax activation is used on the output layer when training. Softmax provides output values between zero and one, whose values from each neuron in a layer sum to unity. We note that as the output of the two output neurons sum to unity, it is also possible to achieve the same result with a single output neuron. Also for the layers of 128 neurons, batch normalisation \citep{2015arXiv150203167I} is applied before ReLU activation, while dropout \citep{JMLR:v15:srivastava14a} is applied after activation in these layers, with a dropout rate of 20\%. All layers are fully connected, that is all the neurons in a layer take all the outputs from the layer below as an input. The FCN has 19\,328 trainable parameters.

\begin{figure*}
	\resizebox{\hsize}{!}{\includegraphics{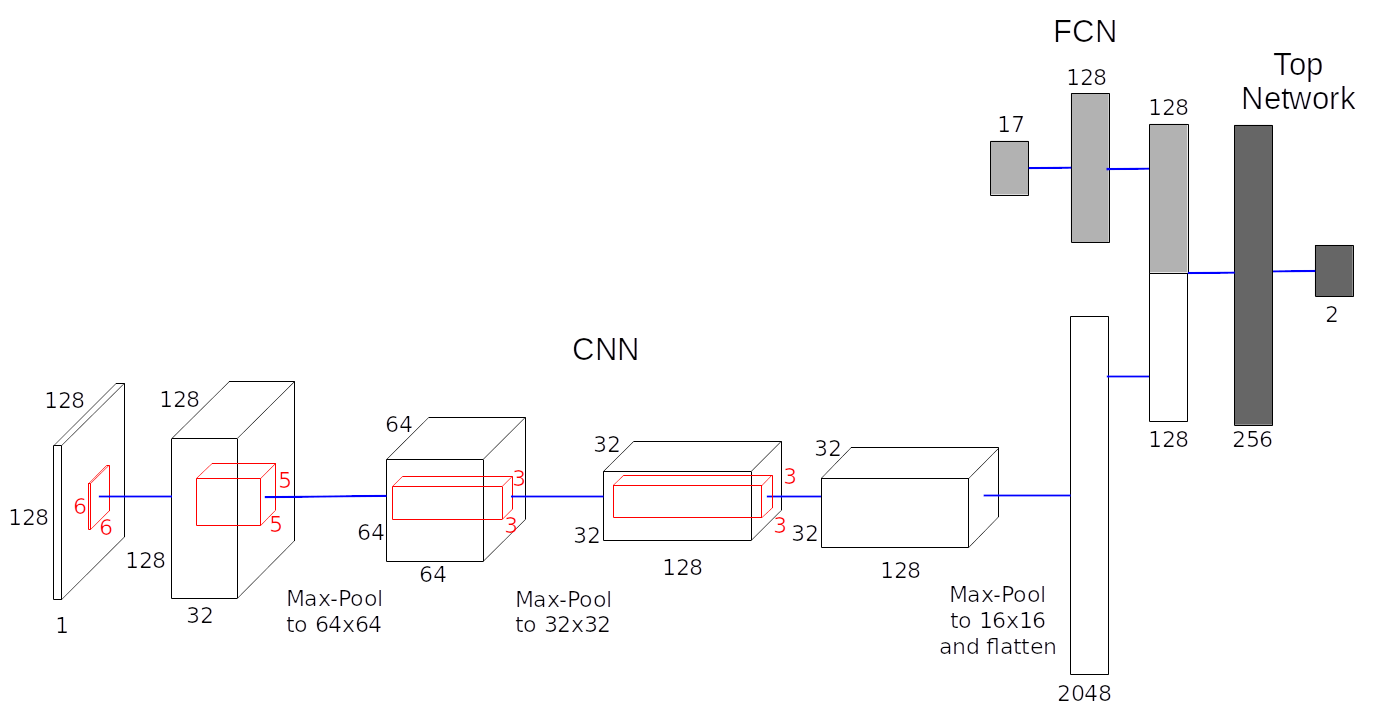}}
	\caption{Visual schematic for the full neural network. White regions are the CNN, light shaded regions are the FCN and the dark shaded regions are the Top Network. The input to the CNN is a single band, 128$\times$128 pixel image while the input to the FCN is the 17 morphological parameters. Both of these inputs are on the left of their corresponding part of the network. The output from the full network is on the right as a two neuron, binary classification. The sizes of the kernels for the CNN (red) and fully connected layers for all parts of the network are shown. The blue line between layers represent the batch normalisation, ReLU activation and dropout that is applied between layers. The CNN has 67\,652\,128 trainable parameters, the FCN has 19\,328 trainable parameters and the Top Network has 66\,818 trainable parameters. The full network has a total of 67\,738\,274 trainable parameters.}
	\label{fig:NN}
\end{figure*}

The architecture of the CNN side is based on the CNN of \citet{2019A&A...631A..51P, 2019A&A...626A..49P}, itself based on the \citet{2015MNRAS.450.1441D} architecture. The lowest four layers, the four layers to the left of the CNN section of Fig. \ref{fig:NN}, are convolutional layers while the top two layers, the right most CNN layers in Fig. \ref{fig:NN}, are fully connected layers. The lowest layer, the left most in Fig \ref{fig:NN}, comprises 32 6$\times$6 kernels, followed by a layer of 64 5$\times$5 kernels and then two layers with 128 3$\times$3 kernels. All convolutional layers use a stride of 1. As with the FCN, batch normalisation is applied before ReLU activation and 20\% dropout is applied after activation for all convolutional layers. After the first, second and fourth convolutional layers, 2$\times$2 max-pooling is performed.

After the convolutional layers, two fully connected layers are used, with 2\,048 and 128 neurons. As with the FCN, batch normalisation is applied before ReLU activation and 20\% dropout is applied after activation for both fully connected layers. For training this part of the full network, the output layer is again composed of two neurons, one for the merger classification and one for the non-merger classification. As with this FCN, softmax activation is used in this layer with no batch normalisation or dropout. The CNN has 67\,652\,128 trainable parameters.

The outputs from the last layers of the FCN and CNN are concatenated to form a single layer of 256 values. These are then passed into the Top Network that comprises a fully connected layer of 256 neurons. As with the FCN and fully connected part of the CNN, batch normalisation is applied before ReLU activation. This is followed by 20\% dropout while training. The output from the Top Network is a layer with two neurons, one each for the merger and non-merger classes, with softmax activation. The Top Network has 66\,818 trainable parameters. The full network can be seen in Fig. \ref{fig:NN} and has a total of 67\,738\,274 trainable parameters.

The CNN, FCN, and Top Network were trained separately (see Sect. \ref{subsec:NNtvt}). For each part of the network, the loss of the network was determined using categorical cross-entropy and was optimised using the Adam algorithm \citep{2014arXiv1412.6980K}. The initial learning rate was $5\times10^{-5}$ for the FCN, $5\times10^{-6}$ for the CNN and $5\times10^{-3}$ for the Top Network. The networks themselves were built using Tensorflow 2.3 \citep{tensorflow2015-whitepaper} and are available on \texttt{GitHub}\footnote{\url{https://github.com/wjpearson/NEP-mergers}} along with the learnt parameters.

For the FCN, a number of different hyper-parameter values were explored. Three layer and four layer architectures were tested, with no improvement over the used two layer structure. Also, 256 and 1\,024 neurons per layer were also tested, again with no improvement over the current architecture. As fewer neurons require less data to effectively train, the smaller size of a two layer network with 128 neurons per layer was chosen.

Few different hyper-parameters were explored for the CNN, as this architecture has been found to perform well in identifying galaxy mergers with data from a number of different surveys \citep{2019A&A...631A..51P, 2019A&A...626A..49P}. However, the number of neurons in the fully connected layers in the CNN were explored, testing 1\,024 and 4\,096 neurons in the left most fully connected layer in Fig. \ref{fig:NN} with no marked improvement to performance.

While testing different architectures for the FCN, the size of the last fully connected layer in the CNN was also changed. As the right most layers of the CNN and FCN (in Fig. \ref{fig:NN}) are the same size, 256 and 1\,024 neurons in this layer were also tested, again showing no change over the architecture used here. The last layers in the FCN and CNN were chosen to be the same size to potentially allow equal weight to be placed on both the morphological parameters and the images. The size of the first layer in the Top Network was matched to the size of the concatenated last layers of the FCN and CNN, and so a layer with 1\,024 and 2\,058 neurons was also tested, again showing no improvement over the current architecture.

For all three networks, different initial learning rates for the Adam optimiser were also tested. Here, rates of $5\times10^{-2}$, $5\times10^{-3}$, $5\times10^{-4}$, $5\times10^{-5}$, $5\times10^{-6}$ and $5\times10^{-7}$ were tested. The best performing initial learning rate, one for each part of the network, was then chosen.

\subsection{Training, validation, and testing}\label{subsec:NNtvt}
The FCN, CNN, and full network were trained independently. The FCN and CNN were trained first using the galaxy morphologies and images, respectively. The two-neuron output layers were then removed and the trained weights and biases of the FCN and CNN were fixed. The top layers of the FCN and CNN were then concatenated and the results passed into the Top Network, which was then trained. The output of the full network, the FCN, CNN and Top Network, was then the prediction for if a galaxy is a merger or not.

The training data described in Sect. \ref{sec:known} were split into three groups. For the $z < 0.15$ networks, 2\,692 galaxies were used to train the network, 338 were used to validate the network as it trained, and a final 336 were used to test the network. For the $0.15 < z < 0.30$ network, 2\,514 were used to train the network, 314 were used to validate the network and 314 were used to test the network. The same galaxy samples were used to train each part of the network.

To train the FCN, the morphological parameters were scaled between zero and one by subtracting the minimum value in Table \ref{tab:morph} and dividing by the range between the minimum and maximum values in Table \ref{tab:morph}. We note that the values presented in Table \ref{tab:morph} do not necessarily directly correspond to the maximum and minimum values of the training data, as seen in Fig. \ref{fig:morph}. The FCN was trained for 5\,000 epochs with the epoch that provided the lowest validation loss being used for training the Top Network and classification. To train the CNN, the images were used. These images were linearly scaled, randomly rotated by 0$^{\circ}$, 90$^{\circ}$, 180$^{\circ}$, or 270$^{\circ}$, randomly flipped vertically, then randomly flipped horizontally as they were passed into the network. CNN are known to not be rotationally invariant \citep[e.g.][]{gong2014multi, 7301273, CHANDRASEKHAR2016426}, while the morphology of a galaxy is independent of rotations in the plane of the sky. Thus this rotation and flipping will help generalisability of the network \citep[e.g.][]{2015MNRAS.450.1441D, 2018ApJ...858..114H}. Redshifts of the galaxies were not used inside the networks as the networks would need to be designed with a specific number of redshifts to be passed into it. As the number of galaxies (background and foreground) within each image will be different, some images will have more galaxies than a specified number and others fewer, it was decided to not include these data. The CNN was trained for 200 epochs with the epoch that provided the lowest validation loss being used for training the Top Network and classification. The Top Network was trained for 1\,000 epochs with the epoch that provided the lowest validation loss being used for classification.

\begin{table}
	\caption[]{Minimum and maximum values used to scale the morphological parameters}
	\label{tab:morph}
	\centering
	\begin{tabular}{ccc}
		\hline
		Parameter & Minimum & Maximum\\
		\hline
		Asymmetry (A) & -4.0 & 4.0\\
		Concentration (C) & 0.0 & 6.0\\
		Deviation (D) & 0.0 & 3.0\\
		Ellipticity asymmetry (Eli A) & 0.0 & 1.0\\
		Ellipticity centroid (Eli Cen) & 0.0 & 1.0\\
		Elongation asymmetry (Elo A) & 1.0 & 8.0\\
		Elongation centroid (Elo Cen) & 1.0 & 8.0\\
		Gini & 0.0 & 1.0\\
		Gini-M$_{20}$ bulge (GMB) & -3.0 & 3.0\\
		Gini-M$_{20}$ merger (GMM) & -1.0 & 1.0\\
		Intensity (I) & 0.0 & 1.0\\
		M$_{20}$ & -4.0 & 0.0\\
		Multimode (M) & 0.0 & 1.0\\
		S\'{e}rsic amplitude (SA) & 0.0 & 200.0\\
		S\'{e}rsic ellipticity (SE) & -6.0 & 3.0\\
		S\'{e}rsic index ($n$) & 0.0 & 50.0\\
		Smoothness (S) & -0.4 & 0.4\\
		\hline
	\end{tabular}
\end{table}

\section{Results}\label{sec:results}
\subsection{Morphological Parameters}
Here we examine the morphological parameters that were used to train the neural networks. As can be seen in Fig. \ref{fig:morph}, some of the derived asymmetries are negative; due to the asymmetry being a sum of residuals, it should always be positive. In theory, the intrinsic asymmetry of a galaxy should be positive but it cannot be measured directly due to the presence of noise. As an attempt to remove the contribution of the background, the corrected asymmetry A$_{corr}$ = A$_{obs}$ - A$_{bkg}$ is typically used \citep{2000ApJ...529..886C}, where A$_{obs}$ is the uncorrected asymmetry and A$_{bkg}$ is the asymmetry of the background. Therefore, negative asymmetries are mathematically allowed as a result of over-correcting for the asymmetry of the background. In general, correctly accounting for noise when measuring the asymmetry parameter is a non-trivial task and is still the topic of active research \citep[e.g.][]{2021MNRAS.507..886T}. Large fractions of negative asymmetry, and smoothness as also seen here, are also found in other observational works \citep[e.g.][]{2019MNRAS.483.4140R, 2020ApJ...899...85S}. Inspection of the positioning of the skyboxes, used to estimate the background noise, and the segmentation maps for a random sample of objects with negative asymmetry did not greatly differ from a random sample of galaxies with positive asymmetry. Thus, we deem the asymmetry and smoothness to be adequate for this work.

We also note that there are negative S\'{e}rsic ellipticities in Fig. \ref{fig:morph} as well as values above unity. While the S\'{e}rsic ellipticity should lie between zero and unity, \texttt{statmorph} allows for fitting to values outside of this range. These can be converted to an equivalent ellipticity (SE$'$) within the range [0,1] using SE$'$~=~min(SE,~2-SE) for SE~$> 1$ or SE$'$~=~max(SE,~SE/(SE-1)) for SE~$< 0$. As these conversions are simple we elected to use the S\'{e}rsic ellipticity from \texttt{statmorph} in this work without conversion. Use of the S\'{e}rsic ellipticity in the presented with this paper in future works should use these conversions.

To check the validity of the morphologies used to train the networks, the morphological parameters of the $z < 0.15$ training data derived from the HSC-SSP images were compared to those derived from GAMA-KiDS data in \citet{2019A&A...631A..51P}. This allows the comparison of the morphologies of the same galaxies using different data. The morphological parameters from the HSC-SSP data were subtracted from those of the GAMA-KiDS data, with the resulting differences ($\delta$) presented in Table \ref{tab:morph-check}. Outliers are defined as galaxies whose morphological parameters are outside of 5$\sigma$ of the mean, where $\sigma$ is the sample standard deviation.

\begin{table}
	\caption[]{Comparison between morphological parameters derived from HSC-SSP images used to train the low redshift network in this work and derived from GAMA-KiDS data in \citet{2019A&A...626A..49P}. These differences are expressed the value from HSC-SSP morphological parameters subtracted from the GAMA-KiDS morphologies ($\delta$): negative means imply that HSC-SSP morphologies are larger than GAMA-KiDS.}
	\label{tab:morph-check}
	\centering
	\begin{tabular}{cccc}
		\hline
		Parameter & Mean $\delta$ & Std $\delta$ & Outliers\\
		\hline
		Asymmetry (A) & -0.060 & 0.087 & 22\\
		Concentration (C) & -0.052 & 0.299 & 26\\
		Deviation (D) & 0.011 & 0.097 & 28\\
		Ellipticity asymmetry (Eli A) & -0.003 & 0.095 & 22\\
		Ellipticity centroid (Eli Cen) & -0.003 & 0.096 & 22\\
		Elongation asymmetry (Elo A) & 0.073 & 2.882 & 4\\
		Elongation centroid (Elo Cen) & 0.051 & 2.694 & 2\\
		Gini & 0.010 & 0.052 & 21\\
		Gini-M$_{20}$ bulge (GMB) & 0.027 & 0.235 & 14\\
		Gini-M$_{20}$ merger (GMM) & 0.013 & 0.068 & 24\\
		Intensity (I) & -0.023 & 0.160 & 38\\
		M$_{20}$ & 0.029 & 0.200 & 28\\
		Multimode (M) & 0.000 & 0.158 & 55\\
		S\'{e}rsic amplitude\tablefootmark{a} (SA) & 0.330 & 1.077 & 21\\
		S\'{e}rsic ellipticity (SE) & -0.044 & 0.150 & 33\\
		S\'{e}rsic index ($n$) & -0.433 & 6.744 & 9\\
		Smoothness (S) & -0.039 & 0.367 & 24\\
		\hline
	\end{tabular}
	\tablefoot{
		\tablefoottext{a}{Comparison is made with surface brightness (mag~arcsec$^{-2}$), not the counts reported in Table \ref{tab:NEP-morph} and \citet{2019A&A...631A..51P}.}
	}
\end{table}

Generally, the results using the HSC-SSP are in good agreement with the morphologies from GAMA-KiDS. For the S\'{e}rsic amplitude, as the photometric zero-points and pixel areas are different for HSC-SSP and KiDS, the comparison is made with the surface brightness in mag~arcsec$^{-2}$ and not the counts, the latter of which are presented in Table \ref{tab:NEP-morph} and \citet{2019A&A...631A..51P}. Thus, the positive mean $\delta$ for SA indicates that the HSC-SSP values are brighter than the GAMA-KiDS values.

None of the resulting distributions are Gaussian, thus we cannot use the expected number of 5$\sigma$ outliers to check the closeness of fit. Chebyshev's inequality restricts the number of objects more than 5$\sigma$ from the mean to be 1/25 of the total number of objects, that is no more than 134 of the 3\,366 galaxies can be classified as outliers.  For all but Multimode and Intensity, there are fewer outliers than a quarter of this value. Multimode and Intensity also have the most non-Gaussian distributions so the higher numbers of outliers may be expected.

Elongation asymmetry, elongation centroid, and S\'{e}rsic index have large standard deviations. For the elongation asymmetry and centroid, these large standard deviations are driven by large values from the GAMA-KiDS morphologies; all outliers have large parameter values compared to the rest of the population. For the S\'{e}rsic index, the large standard deviation is driven by a small number of galaxies with a large S\'{e}rsic index in either the HSC-SSP data or GAMA-KiDS, with four out of nine of the outliers being due to large $n$ in GAMA-KiDS and five being due to large HSC-SSP $n$.

\subsection{Galaxy mergers}
\begin{figure}
	\resizebox{\hsize}{!}{\includegraphics{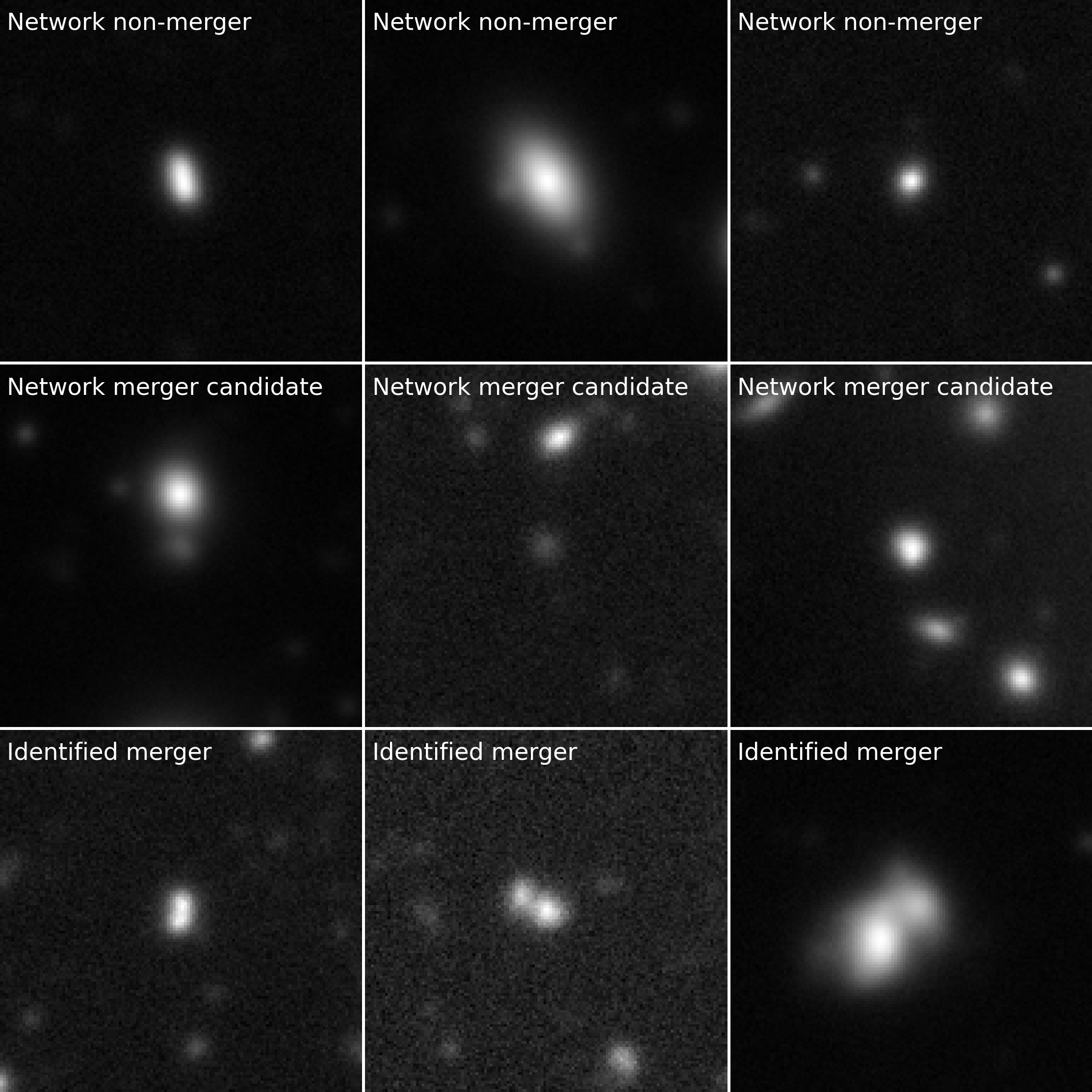}}
	\caption{Randomly selected HSC-NEP galaxies as example for galaxies selected by the networks as non-mergers (top row), galaxies selected as merger candidates by the networks but not identified as mergers by visual inspection (middle row), and galaxies selected as galaxy mergers by visual inspection (bottom row). The classified galaxy lies in the centre of the 128$\times$128 pixel ($\approx$ 21.5$\times$21.5 arcsec) image and the images are shown with asinh scaling.}
	\label{fig:examples}
\end{figure}

In this section, we present the results of our model's test performance and outline our visual inspection programme. An example of the final catalogue is presented in Table \ref{tab:cat}. The results from the neural networks are given as the probability that a galaxy is a merger or non-merger, \texttt{frac\_merger} and \texttt{frac\_nonmerger} respectively. It also has the classification from visual inspection as \texttt{vis\_merger} (see Sect. \ref{subsec:vis} below). Randomly selected examples of HSC-NEP galaxies identified as non-mergers by the networks, as mergers by the networks but not visual inspection, and as mergers by visual inspection are presented in Fig. \ref{fig:examples}. Here, we take galaxies with \texttt{frac\_merger} $> 0.5$ to be identified as mergers by the networks (hereafter merger candidates).

\begin{table*}
	\caption[]{Ten rows of the catalogue of merging galaxies in NEP}
	\label{tab:cat}
	\centering
	\begin{tabular}{cccccc}
		\hline
		HSC\_ID & RA & Dec & frac\_merger & frac\_nonmerger & vis\_merger\\
		\hline
		79217643822780147 & 270.743 & 65.328 & 0.572 & 0.428 & False\\
		79217643822780325 & 270.782 & 65.341 & 0.988 & 0.012 & False\\
		79217643822780333 & 270.789 & 65.343 & 0.025 & 0.975 & False\\
		79217643822780337 & 270.795 & 65.340 & 0.056 & 0.944 & False\\
		79217643822781103 & 270.745 & 65.346 & 0.088 & 0.912 & False\\
		79217648117743947 & 270.775 & 65.386 & 0.182 & 0.818 & False\\
		79217648117753062 & 270.734 & 65.374 & 0.056 & 0.944 & False\\
		79217648117753093 & 270.758 & 65.361 & 0.145 & 0.855 & False\\
		79217648117753463 & 270.737 & 65.374 & 0.286 & 0.714 & False\\
		79666772847897740 & 267.434 & 65.748 & 0.959 & 0.041 & True\\
		... & ... & ... & ... & ... & ...\\
		\hline
	\end{tabular}
\end{table*}

\subsubsection{Neural networks}\label{subsec:nn}
In determining the architecture of the neural network, it was found that combining a FCN and CNN had better performance than a FCN or CNN alone. Tests with the $z < 0.15$ data set, the validation of the best FCN had a loss of 0.301 and accuracy of 88.8\% while the validation of the best CNN had a loss of 0.473 and accuracy of 79.3\%. When combining the FCN and CNN, as described in Sect. \ref{subsec:NNa}, the validation of the final full network for $z < 0.15$ galaxies has a loss of 0.260 and accuracy of 91.7\%. It would be expected that there is information contained in the images that is not present in the morphological parameters: the morphological parameters can be seen as a compression of the information of the images. However, this result suggests that there is information in the morphological parameters that is not present in the images, or more likely the information in the morphological parameters is more easily extracted by a neural network than the information in the images. This difficulty may lie in the noise or background of the image confusing the network. The same noise or background may present a similar issue for the morphological parameter extraction but the network itself is presented the pre-extracted parameters. Thus, combining the images and morphologies allows the network to supplement the more easily interpreted morphological parameters with further, harder to extract information contained within the images. As a result, it is not entirely surprising that the network performs better combining the images with the morphological parameters than either alone despite both containing similar information.

The quality of the two full networks, one for $z < 0.15$ and one for $0.15 \leq z < 0.30$, can be determined by the results presented in Table \ref{tab:nw}. Due to the training set being class balanced while mergers are expected to be in the minority of real galaxies, we caution the use of accuracy alone to determine the quality of a network when applied to non-class balanced data.

The trained networks described in Sects. \ref{subsec:NNa} and \ref{subsec:NNtvt} were applied to galaxies in the North Ecliptic Pole. Taking a galaxy with \texttt{frac\_merger} greater than 0.5 as a merger candidate, these classifications resulted in 1\,477 of 6\,965 galaxies at $z < 0.15$ and 8\,718 of 27\,299 galaxies at $0.15 \leq z < 0.30$ being identified as galaxy merger candidates. This results in a merger candidate fraction of 21.2\% for the lower redshift range and 31.9\% for the higher redshift range.

\begin{table}
	\caption[]{Performance statistics from the neural networks. All values calculated with the class balanced test data set.}
	\label{tab:nw}
	\centering
	\begin{tabular}{lcc}
		\hline
		Redshift & Statistic & Value\\
		\hline
		\multirow{5}{*}{$z < 0.15$} & Accuracy & 0.884\\
		 & Recall & 0.863\\
		 & Precision & 0.901\\
		 & Specificity & 0.905\\
		 & NPV\tablefootmark{a} & 0.869\\
		\hline
		\multirow{5}{*}{$0.15 \leq z < 0.30$} & Accuracy & 0.850\\
		 & Recall & 0.790\\
		 & Precision & 0.899\\
		 & Specificity & 0.911\\
		 & NPV\tablefootmark{a} & 0.812\\
		\hline
	\end{tabular}
	\tablefoot{
		\tablefoottext{a}{Negative predictive value}\\
		Defenitions of the statistics can be found in Appendix \ref{app:definitions}.
	}
\end{table}

\subsubsection{Visual Inspection}\label{subsec:vis}
As we expect there to be a large number of falsely identified galaxy mergers in the merger candidates identified by our full networks, the galaxies identified as galaxy merger candidates by the full network were visually checked by two authors, the majority by WJP and a minority by LES. Discussion of the quality of the visual classifiers can be found in Appendix \ref{app:authors}. The visual classification includes considering the redshifts of galaxies close to the merger candidate to check for close companions. If both WJP and LES inspected a merger candidate, a galaxy was only considered a merger if both WJP and LES considered the galaxy to be a merger. This resulted in 251 of 1\,477 being true mergers at $z < 0.15$ and 1858 of 8718 being true mergers at $0.15 \leq z < 0.30$. This results in a merger fraction of $3.6\%$ for $z < 0.15$ and $6.8\%$ for $0.15 \leq z < 0.30$. However, due to the difficulties in visual classification, it is possible that some of the galaxies identified as merger candidates by the networks could truly be mergers but misclassified as non-mergers during visual inspection.

With the large number of non-merging galaxies that would need to be visually checked, it was deemed too time costly to visually confirm all non-mergers. As the number of mergers is expected to be low and the recall of the network is high, very few true mergers (approximately 13.7\% at $z < 0.15$ and 21.0\% at $0.15 \leq z < 0.30$) are expected to be classified as non-mergers and so few mergers are expected to be missed. Using the recall of the two full networks presented in Table \ref{tab:nw} and the number of visually confirmed mergers, we expect to miss approximately 40 mergers at $z < 0.15$ and approximately 494 at $0.15 \leq z < 0.30$.

However, as discussed in Appendix \ref{app:authors}, the visual classifications are not complete with a recall of 0.45. If we combine this with the network performances presented in Table \ref{tab:nw}, we expect the final visually selected merger samples to be 38.8\% complete at $z < 0.15$ and 35.6\% complete at $0.15 \leq z < 0.30$. The test merger candidate samples contain 9.5\% and 8.9\% of all non-mergers at $z < 0.15$ and $0.15 \leq z < 0.30$, respectively. Again combining these with the average specificity of the visual classifiers, the visually selected merger samples contain 1.9\% and 1.8\% of all non-mergers at $z < 0.15$ and $0.15 \leq z < 0.30$, respectively. If we take the true merger fractions to be 3.6\% at $z < 0.15$ and 6.8\% at $0.15 \leq z < 0.30$, this implies the visually confirmed merger samples are 43.3\% pure at $z < 0.15$ and 59.0\% pure at $0.15 \leq z < 0.30$. However, as the visual classification was done on a pre-selected sample of merger candidates while the discussion in Appendix \ref{app:authors} was performed with a class balanced sample of mergers and non-mergers with no pre-selection on morphologies, the quality of the visual classifiers may be lower than presented, a result of a pre-selected sample likely being harder to differentiate between mergers and non-mergers than an unselected sample.

We also visually inspected a small sample of galaxies identified as non-mergers, 100 from the $z < 0.15$ network and 100 from the $0.15 \leq z < 0.30$ network. Within both of these samples we found no obvious misclassifications, supporting the expectation that very few mergers were misclassified as non-mergers by the networks. A summary of the number of galaxies identified as non-merger, merger candidates and visually selected mergers is presented in Table \ref{tab:summary}.

\begin{table}
	\caption[]{Summary of the number of galaxies identified as non-mergers, merger candidates and visually confirmed mergers}
	\label{tab:summary}
	\centering
	\setlength{\tabcolsep}{3pt}
	\begin{tabular}{ccccc}
		\hline
		& \multirow{2}{*}{\shortstack{Total\\galaxies}} & & \multirow{2}{*}{\shortstack{Merger\\candidate}} & \multirow{2}{*}{\shortstack{Confirmed\\merger}}\\
		Redshift & & Non-merger & &\\
		\hline
		$z < 0.15$ & 6\,965 & 5\,488 & 1\,477 & 251\\
		$0.15 \leq z < 0.30$ & 27\,299 & 18\,581 & 8\,718 & 1\,858\\
		\hline
	\end{tabular}
\end{table}

During visual inspection, the non-mergers were also briefly checked for true non-merging galaxies with visible structure. This is due to the training sample possibly causing the networks to be trained on structureless and structured galaxies, and not mergers and non-mergers, as discussed in Sect. \ref{sec:known}. The brief visual inspection found that there were galaxies identified as non-mergers that did contain resolvable structure, such as spiral arms, as can be seen in Fig. \ref{fig:structure}. Similarly, there are merger candidate galaxies that have no visible structure, as shown in Appendix \ref{app:random}. Thus, the networks have not been inadvertently trained to find structured and non-structured galaxies.

\begin{figure}
	\resizebox{\hsize}{!}{\includegraphics{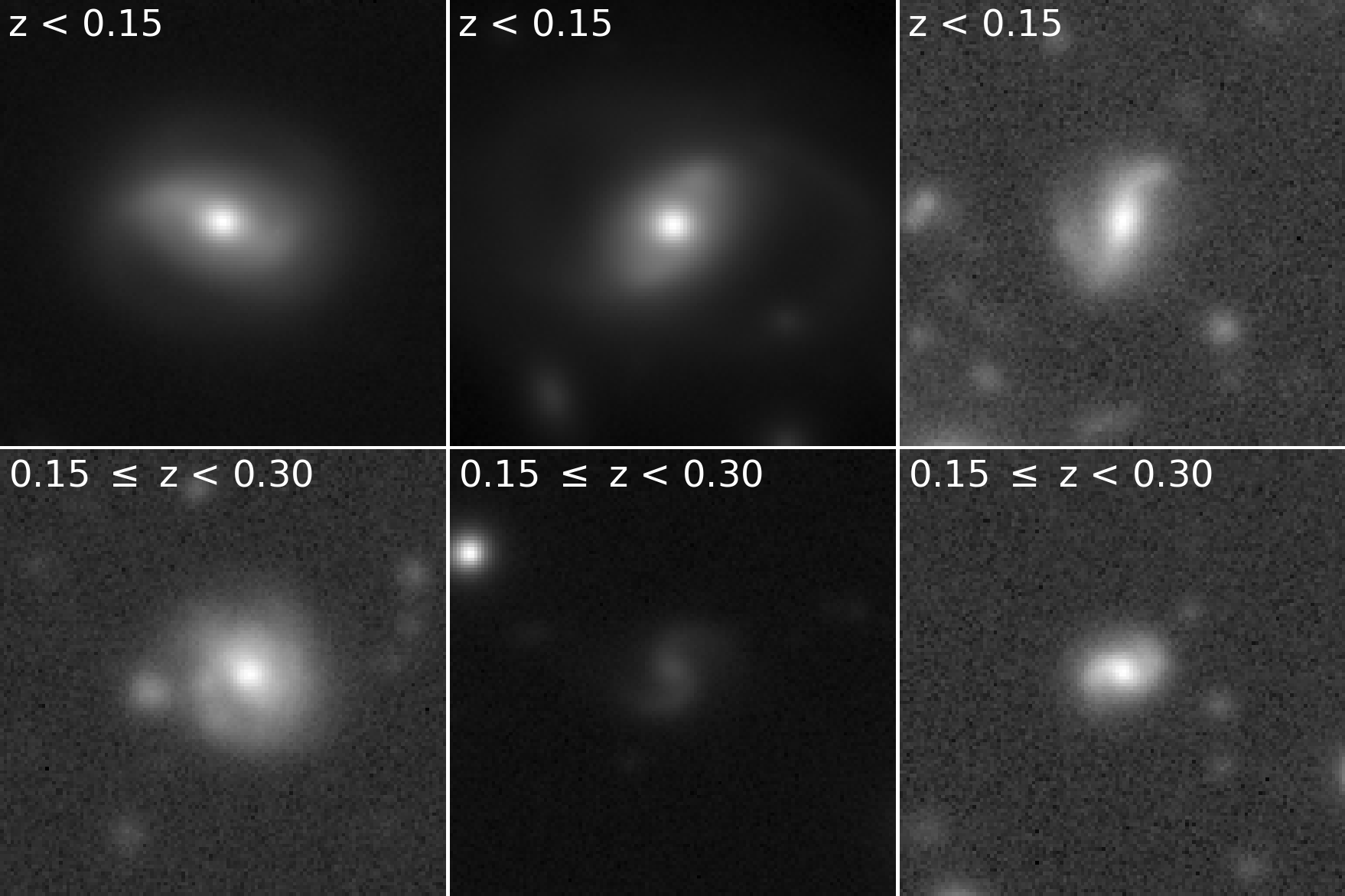}}
	\caption{Examples of visually confirmed non-mergers with visible structure. The top row shows non-mergers detected by the $z < 0.15$ network and the bottom row shows non-mergers detected by the $0.15 \leq z < 0.30$ network.}
	\label{fig:structure}
\end{figure}

\section{Discussion}\label{sec:discuss}
\subsection{False positives}
While it is expected that there will be false positive (FP) detections from the networks, that is galaxies that are identified by the full network as a merger which are non-mergers, it is informative to understand why such galaxies are misclassified.

\subsubsection{Image Occlusion}
\begin{figure*}
	\resizebox{\hsize}{!}{\includegraphics{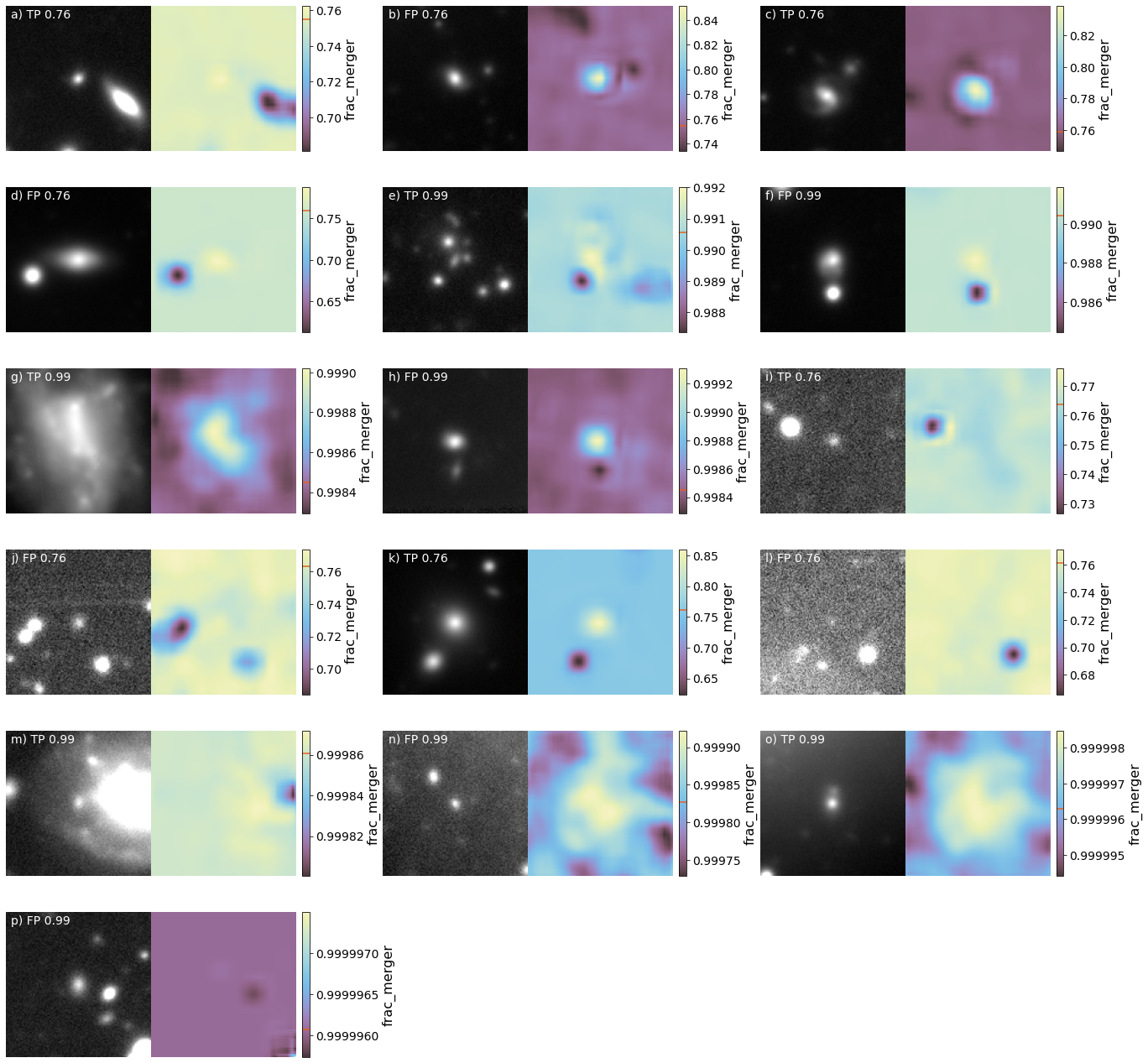}}
	\caption{Galaxy images and associated feature importance heatmaps for sixteen galaxies at $z < 0.15$. The sixteen galaxies comprise of four TP galaxies with \texttt{frac\_merger}$\approx$0.76, four TP galaxies with \texttt{frac\_merger}$\approx$0.99, four FP galaxies with \texttt{frac\_merger}$\approx$0.76, and four FP galaxies with \texttt{frac\_merger}$\approx$0.99, as indicated in the upper left corners of the galaxy images. The panels to the right of the galaxy images contain the heat maps for feature importance, with pixels that when obscured cause the galaxy to have lower \texttt{frac\_merger} in dark purple and pixels that cause the galaxy to have a higher \texttt{frac\_merger} in light yellow. The orange line in the colour bar indicates the \texttt{frac\_merger} of the un-occluded galaxy.}
	\label{fig:occ-0p0}
\end{figure*}

To understand which visual properties of the FP galaxies were being identified for galaxies at $z < 0.15$, occlusion experiments \citep[e.g.][]{Zeiler-Fergus-2014, ancona2018towards, 2019A&A...626A..49P, 2020A&A...644A..87W} were performed on four FP galaxies with \texttt{frac\_merger}$\approx$0.76 (galaxies b, d, j, and l in Fig. \ref{fig:occ-0p0}) and four with \texttt{frac\_merger}$\approx$0.99 (galaxies f, h, n, and p in Fig. \ref{fig:occ-0p0}). Eight true positive (TP) galaxies were also selected for occlusion experiments: four with \texttt{frac\_merger}$\approx$0.76 (galaxies a, c, i, and k) and four with \texttt{frac\_merger}$\approx$0.99 (galaxies e, g, m, and o). For this experiment, a $16 \times 16$ pixel region of the images were set to zero. The $16 \times 16$ pixel zero region was translated across the image by one pixel such that there were a total of 12\,769 copies of the galaxy with a different $16 \times 16$ pixel regions set to zero. These occluded images were then passed through the full network with the morphological parameters left unchanged. The occluded galaxy images are treated as a normal galaxy by the networks and so are scaled by the networks to be between zero, the faintest pixel in the occluded image, and one, the brightest pixel in the occluded image. Heat maps were then generated by taking the average classification for when each pixel was occluded. Figure \ref{fig:occ-0p0} shows these heat-maps along with the original image of the galaxy.

\begin{figure*}
	\resizebox{\hsize}{!}{\includegraphics{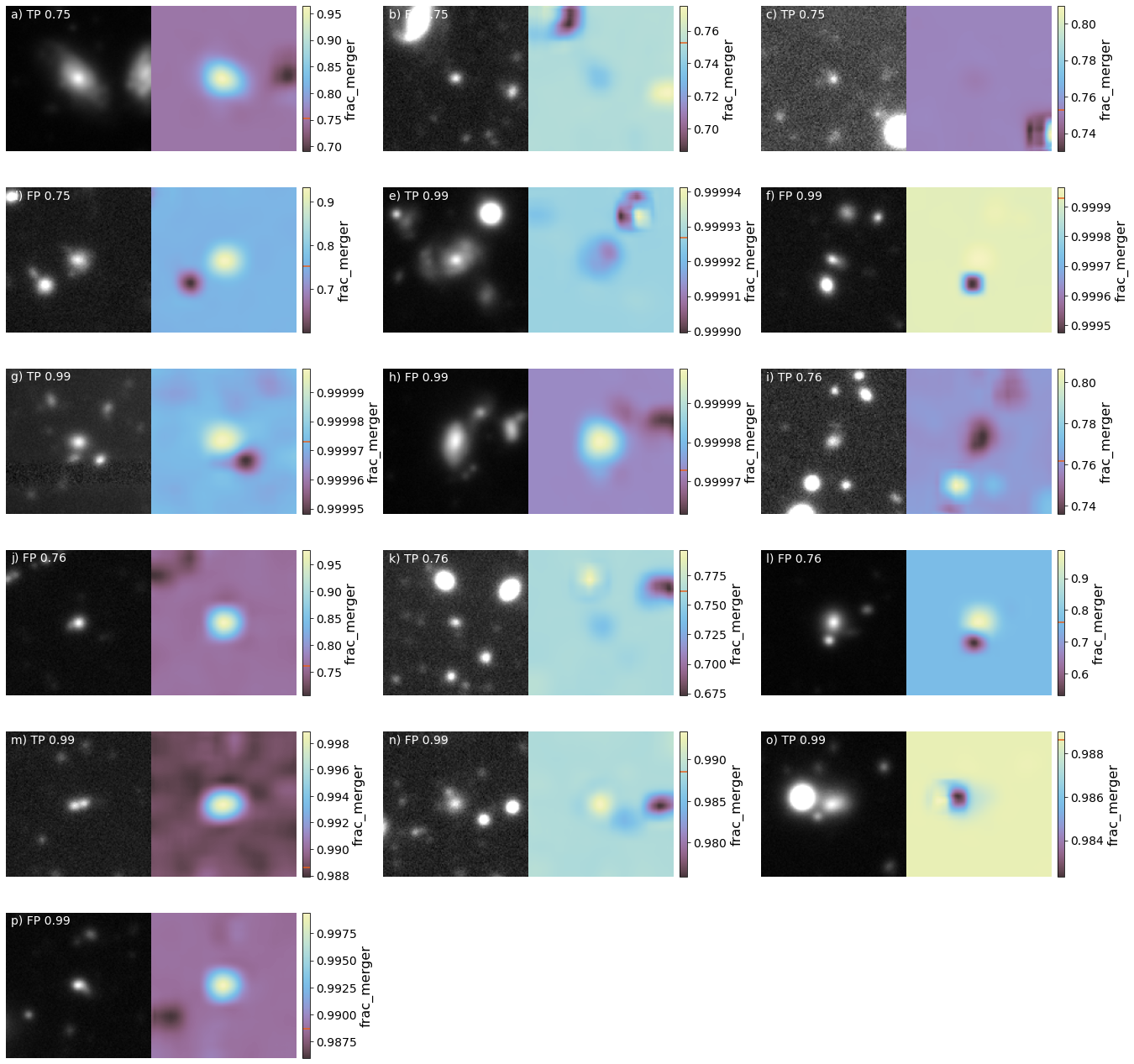}}
	\caption{Same as Fig. \ref{fig:occ-0p0} but for the $0.15 \leq z < 0.30$ network. Galaxy images and associated feature importance heatmaps for sixteen galaxies at $0.15 \leq z < 0.30$. The sixteen galaxies comprise of four TP galaxies with \texttt{frac\_merger}$\approx$0.76, four TP galaxies with \texttt{frac\_merger}$\approx$0.99, four FP galaxies with \texttt{frac\_merger}$\approx$0.76, and four FP galaxies with \texttt{frac\_merger}$\approx$0.99, as indicated in the upper left corners of the galaxy images. The panels to the right of the galaxy images contain the heat maps for feature importance, with pixels that when obscured cause the galaxy to have lower \texttt{frac\_merger} in dark purple and pixels that cause the galaxy to have a higher \texttt{frac\_merger} in light yellow. The orange line in the colour bar indicates the \texttt{frac\_merger} of the un-occluded galaxy.}
	\label{fig:occ-0p15}
\end{figure*}

The heat-maps in Figs. \ref{fig:occ-0p0}a and \ref{fig:occ-0p15} indicate the regions that are important for the CNN part of the network to identify a merging galaxy. Each pixel within these images indicates the average change of classification when the pixel is occluded. As the average is of up to 256 values, large changes when the pixel is occluded will be suppressed. This means that Figs. \ref{fig:occ-0p0} and \ref{fig:occ-0p15} are primarily useful for qualitative analysis. Thus, while no galaxies seen in these figures show a change in classification and suggest the classification is primarily driven by the morphologies, these plots cannot be used for such definitive statements.

For all FPs, the presence of the second galaxy in the frame is an important component used for classification. These secondary galaxies are not physically associated with the primary galaxies in the centre of the image due to their different redshifts. Occlusion of these secondary galaxies reveals that the full network is interpreting them as potential merging companions.As the redshift information is not passed into the network, this is a somewhat understandable mistake. However, the weak reliance on the images by the full network means the presence of the secondary galaxy in the image is not of great importance overall.

The secondary galaxy influencing classification is also seen with the TPs (a), (e), (i), (k), and (m). For the remaining TPs, instead of being influenced by a secondary galaxy the network is identifying faint features around the primary galaxy, likely signatures of tidal disruption.

From the comparison of the FP and TP, there is the suggestion that including the redshift of the primary and secondary galaxies may aid in determining if two galaxies are indeed merging or are just close in projection but are not physically associated. This was not done due to the reasons previously outlined in Sect. \ref{subsec:NNtvt}.

The majority of galaxies show that the primary galaxy is also used in determining the classification. Only galaxies (i), (j), (l), (m), and (p) do not show this behaviour. It is unclear why obscuring the primary galaxy makes a galaxy more likely to be seen to contain a merger. Hiding of the central source may make fainter structures around the galaxy more apparent and hence easier to identify as a merger, but this is speculation.

For the higher redshift network, the image occlusion provides similar results, as seen in Fig. \ref{fig:occ-0p15}. All FP galaxies show the presence of a secondary galaxy is important for classification, with an apparent reduction in\texttt{frac\_merger} when it is abscured. Only the TP (i) and (m) galaxies do not see a reduction when a secondary galaxy is obscured. In the case of (i), the merging galaxies are very close to one another making obscuration of a single galaxy of the pair difficult. The high redshift network also sees an influence to classification when the primary galaxy is obscured for galaxies (a), (d), (g), (h), (j), (l), (m), (n), and (p), similar to the low redshift network.

However, none of the sixteen, higher redshift galaxies that had the occlusion experiment performed show the importance of faint structures. This does not mean that such structures are not important to the network, just that such structures are not important for the sixteen galaxies shown. Galaxy (i) also exhibits occlusion behaviour that is opposite to what is seen in all other galaxies at both redshifts. For galaxy (i), the occlusion of the primary galaxy reduces \texttt{frac\_merger} while the occlusion of the bright object in the field of view increases \texttt{frac\_merger}.

\begin{figure}
	\resizebox{\hsize}{!}{\includegraphics{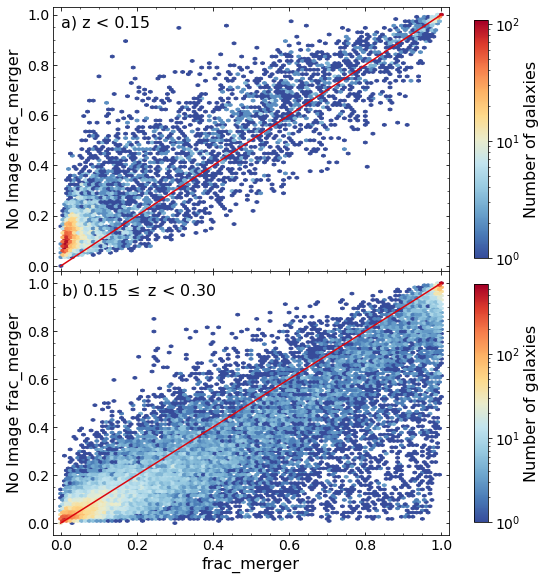}}
	\caption{\texttt{frac\_merger} when the image is occluded as a function of the original \texttt{frac\_merger} for all galaxies classified by the low redshift network (panel a) and the high redshift network (panel b). Number density of galaxies is shown from low (dark blue) to high (red). The  1-to-1 line is shown in red.}
	\label{fig:noImg}
\end{figure}

We also fully occluded all galaxies, that is we passed an array of zeros in place of the image into the network, and compared the resulting \texttt{frac\_merger} with the original classification. As can be seen in Fig. \ref{fig:noImg}a, the low redshift network's new classifications are typically slightly higher for the fully occluded images at lower \texttt{frac\_merger} before becoming consistent at higher \texttt{frac\_merger}. There is a positive correlation between the two classifications, although with a large scatter of approximately 0.1. This suggests that, while useful in determining classification, the images are not a strong influence on the classification when compared to the morphologies being fully occluded (Sect. \ref{subsec:morphocc}).

For the high redshift network, there is good agreement between the original \texttt{frac\_merger} and the image occluded \texttt{frac\_merger} at low \texttt{frac\_merger}. As \texttt{frac\_merger} increases, the occluded \texttt{frac\_merger} typically has a lower value, as can be seen in Fig. \ref{fig:noImg}b. There is a large number of objects with the occluded \texttt{frac\_merger} close to zero while the un-occluded \texttt{frac\_merger} is much larger, a trend not seen in the lower network. This suggests that the images have more importance for the classification than the lower redshift network. However, the correlation between the original \texttt{frac\_merger} and the occluded \texttt{frac\_merger} suggests that the images still play a minor roll in classification. The morphological parameter occlusion discussed below is in support of the minor importance of the images for the higher redshift network.

\subsubsection{Morphological Parameter Occlusion}\label{subsec:morphocc}
\begin{figure*}
	\centering
	\resizebox{0.81\hsize}{!}{\includegraphics{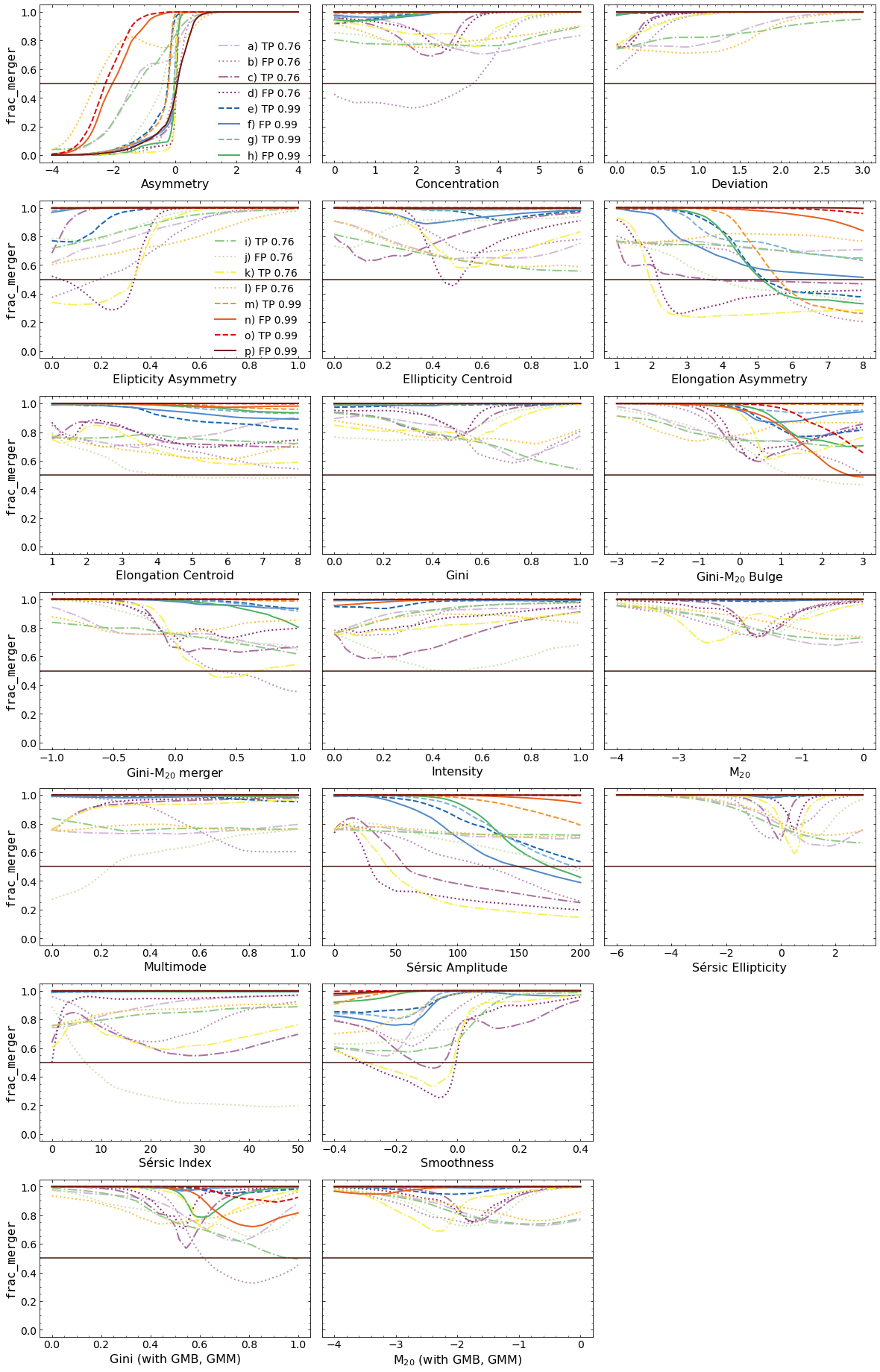}}
	\caption{Change in \texttt{frac\_merger} due to change in the morphological parameters for sixteen galaxies at $z < 0.15$: four TP galaxies with \texttt{frac\_merger}$\approx$0.76 (dot-dashed lines), four TP galaxies with \texttt{frac\_merger}$\approx$0.99 (dashed lines), four FP galaxies with \texttt{frac\_merger}$\approx$0.76 (dotted lines), and four FP galaxies with \texttt{frac\_merger}$\approx$0.99 (solid lines). The horizontal line indicates the decision threshold with the merger class being above and the non-merger class being below the line. The sixteen galaxies correspond to the sixteen galaxies in Fig. \ref{fig:occ-0p0}.}
	\label{fig:morph-occ-0p0}
\end{figure*}

\begin{figure*}
	\centering
	\resizebox{0.81\hsize}{!}{\includegraphics{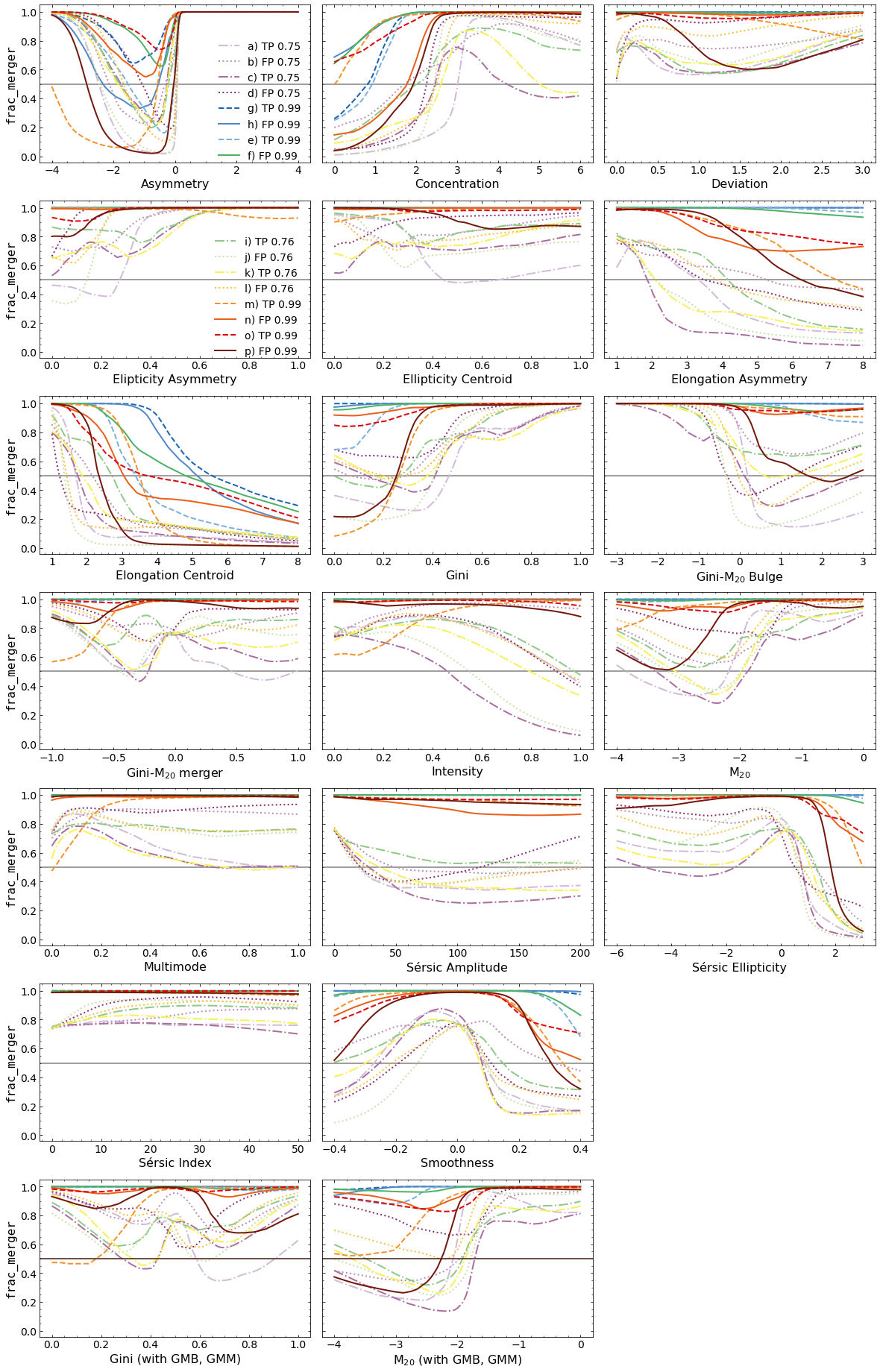}}
	\caption{Same as Fig. \ref{fig:morph-occ-0p0} but for the $0.15 \leq z < 0.30$ network. Change in \texttt{frac\_merger} due to change in the morphological parameters for sixteen galaxies: four TP galaxies with \texttt{frac\_merger}$\approx$0.76 (dot-dashed lines), four TP galaxies with \texttt{frac\_merger}$\approx$0.99 (dashed lines), four FP galaxies with \texttt{frac\_merger}$\approx$0.76 (dotted lines), and four FP galaxies with \texttt{frac\_merger}$\approx$0.99 (solid lines). The horizontal line indicates the decision threshold with the merger class being above and the non-merger class being below the line. The sixteen galaxies correspond to the sixteen galaxies in Fig. \ref{fig:occ-0p15}.}
	\label{fig:morph-occ-0p15}
\end{figure*}

Occlusion experiments similar to those applied to the images are difficult to perform with the morphological parameters. The lowest input morphological value into the full network is zero by design (see Sect. \ref{subsec:NNtvt}) and passing negative values could result in unpredictable and non-interpretable behaviour. Instead of setting each morphological parameter to zero, we instead change the value of each parameter. The range of each parameter in Table \ref{tab:morph} was split into 802 equally spaced steps. For each of the 32 galaxies in Figs. \ref{fig:occ-0p0} and \ref{fig:occ-0p15}, each morphological parameter was set to each of these 802 values one at a time. For example, the asymmetry was set to -4 while all other parameters and the image were left alone. As GMB and GMM are linear combinations of Gini and M$_{20}$, we also alter Gini or M$_{20}$ as described above and perform the corresponding change to GMB and GMM following Eqns. \ref{eqn:gmb} and \ref{eqn:gmm}, respectively. These are presented in Figs. \ref{fig:morph-occ-0p0} and \ref{fig:morph-occ-0p15} as `Gini (with GMB, GMM)' and `M$_{20}$ (with GMB, GMM)'. While altering Gini, M$_{20}$, GMB or GMM individually is not representative of real world applications, we make these comparisons for completeness. These galaxies with modified morphological parameters were then classified by the full network so the change in classification as each parameter is changed can be studied. The resulting changes in \texttt{frac\_merger} as the morphological parameters are changed are shown in Fig. \ref{fig:morph-occ-0p0} for the $z < 0.15$ network and Fig. \ref{fig:morph-occ-0p15} for the $0.15 \leq z < 0.30$ network.

Changing the morphological parameters alters \texttt{frac\_merger} in all cases. However, changes in D, Gini, M$_{20}$, I, SE and M$_{20}$ (with GMB, GMM) do not result in a change of classification for any of the 16 $z < 0.15$ galaxies studied. Thus, these parameters are the least important in this network for determining the classification of the galaxy. For a further five parameters, C, Eli Cen, Elo Cen, M, and $n$, only one of the sixteen galaxies sees a change in classification, again indicating that these parameters play a minor role in classification for the $z < 0.15$ network. For these parameters, the galaxies that see a change in classification are all FP with \texttt{frac\_merger}$\approx$0.76.

While Gini and M$_{20}$ are often used to identify galaxy mergers \citep[e.g.][]{2004AJ....128..163L, 2008ApJ...672..177L}, as the training sample was not selected using these parameters it is perhaps not surprising that these two parameters have little importnace. This is also not due to the presence of linear combinations of Gini and M$_{20}$ in GMB and GMM. When GMB and GMM are changed along with Gini or M$_{20}$ as per their definitions, changing M$_{20}$ (with GMB, GMM) does not result in a change in classification for any of the sixteen galaxies while changing Gini (with GMB, GMM) only sees a change in classification for two of the galaxies.

In the other extreme, only changing A changed the classification of all sixteen galaxies at $z < 0.15$, indicating that this is a powerful morphological parameter for identifying merging systems. The Elo A also sees changes in classifications for half of the galaxies studied in detail, further indicating the importance of an asymmetric light distribution in identifying merging galaxies. The Eli A, however, sees changes for fewer galaxies: only three of the sixteen galaxies see a change to classification.

For the remaining parameters for the $z < 0.15$ galaxies, the SA sees a change in classification for half of the sixteen galaxies, indicating that it is an important parameter for this network. The S parameter sees a change in the classification for two TP \texttt{frac\_merger}$\approx$0.76 galaxies and one FP \texttt{frac\_merger}$\approx$0.76 galaxy. The GMB shows a change in classification for one FP \texttt{frac\_merger}$\approx$0.76 galaxy and one FP \texttt{frac\_merger}$\approx$0.99 galaxy, while GMM sees a change in classification for two galaxies. We reiterate that changing GMB or GMM independently of Gini or M$_{20}$ is not representative of the real world and so limited understanding can be gained from changing these two parameters in isolation.

For the higher redshift network, only changes in D and $n$ do not result in a classification change for all sixteen galaxies. In the other extreme, only changing the Elo Cen changes the classification for the $0.15 \leq z < 0.30$ galaxies studied. Like the $z < 0.15$ network, asymmetry is again important for classification for the $0.15 \leq z < 0.30$ network, with only two FP \texttt{frac\_merger}$\approx$0.99 and two TP \texttt{frac\_merger}$\approx$0.99 galaxies not showing a change in classification. Elo A is again shows a change in classification, here for ten of the galaxies. The galaxies that do not see change to the classification all have \texttt{frac\_merger}$\approx$0.99. This again highlights the importance of an asymmetric light distribution in identifying merging galaxies.

Concentration is more important for the higher redshift network than the lower redshift network, with only two FP \texttt{frac\_merger}$\approx$0.99 and two TP \texttt{frac\_merger}$\approx$0.99 galaxies seeing no change in classification. Gini and M$_{20}$ are also more important in the higher redshift network than the lower redshift network, with Gini causing a change in classification to nine galaxies and M$_{20}$ causing a change to four galaxies. GMM and GMB also cause changes to classifications in more galaxies in the higher redshift network than the lower redshift network. Again, changing these four parameters in isolation is not realistic. Changing Gini and M$_{20}$ with GMB and GMM also shows a greater influence on the classification than the lower redshift network. Gini (with GMB, GMM) sees a change in classification for five of the sixteen galaxies while M$_{20}$ (with GMB, GMM) sees a change for seven of the galaxies. This again indicates the stronger reliance on Gini and M$_{20}$ for the higher redsift network compared to the lower redshift network.

The changes in \texttt{frac\_merger} for the morphological parameters were much larger than seen in the occlusion experiments. This supports the idea that the morphological parameters are more important to the full networks than the images for both the $z < 0.15$ and $ 0.15 \leq z < 0.30$ networks. Generally, the higher redshift network appears to rely on a number of different parameters for the classification of galaxies while the lower redshift network primarily sees changes for the parameters that measure the asymmetry of the light distribution. We note caution, however, as these examinations have only been conducted with a small number of galaxies.

\begin{figure}
	\resizebox{\hsize}{!}{\includegraphics{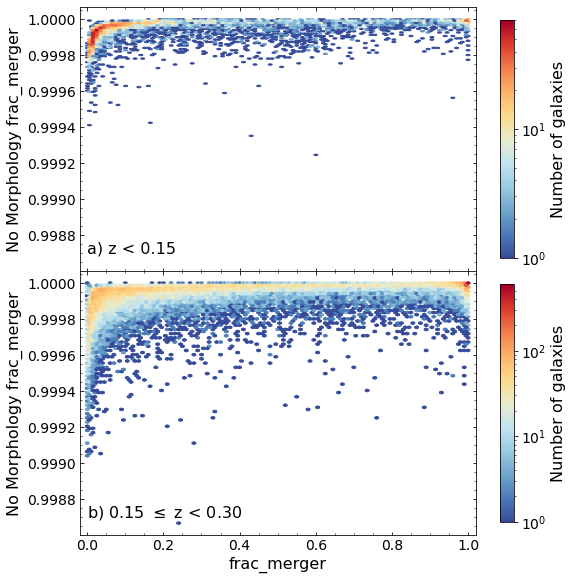}}
	\caption{\texttt{frac\_merger} when all morphological parameters are occluded as a function of the original \texttt{frac\_merger} for all galaxies classified by the low redshift network (panel a) and the high redshift network (panel b). Number density of galaxies is shown from low (dark blue) to high (red).}
	\label{fig:noMph}
\end{figure}

As with the images, we have also occluded the morphologies for all galaxies by setting each morphological parameter to the minimum value in Table \ref{tab:morph} in place of the correct parameter value. For the low redshift network, this sets the \texttt{frac\_merger} for all galaxies close to unity, as can be seen in Fig. \ref{fig:noMph}a. As the image occlusion resulted in a changed but correlated new \texttt{frac\_merger} value, it is apparent that the morphology is the main component used for classification of the galaxies as occluding the morphology has a much larger impact on the resulting \texttt{frac\_merger}. If the images provided no information for classification, then all the galaxies would all have the same \texttt{frac\_merger} when the morphologies are occluded.

A similar trend is seen with the high redshift network. When the morphological parameters are to the minimum value in Table \ref{tab:morph}, the \texttt{frac\_merger} of all galaxies becomes close to unity, as can be seen in Fig \ref{fig:noMph}b. Again, the large change in \texttt{frac\_merger} when the morphologies are occluded while the changes to \texttt{frac\_merger} due to image occlusion are not as severe implies that the high redshift network is primarily using information from the morphologies to determine the classification.

\subsection{Merger fraction}\label{subsec:mgr-frac}
\begin{figure}
	\resizebox{\hsize}{!}{\includegraphics{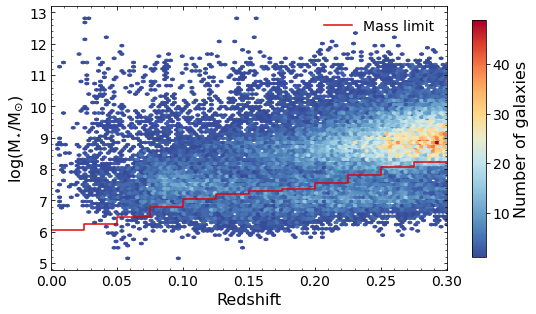}}
	\caption{Stellar mass as a function of redshift. The number of galaxies in each mass-redshift bin is from low in blue to high in red. The mass limits calculated following Sect. \ref{subsec:mass} are shown as a red line.}
	\label{fig:masslim}
\end{figure}

\begin{figure}
	\resizebox{\hsize}{!}{\includegraphics{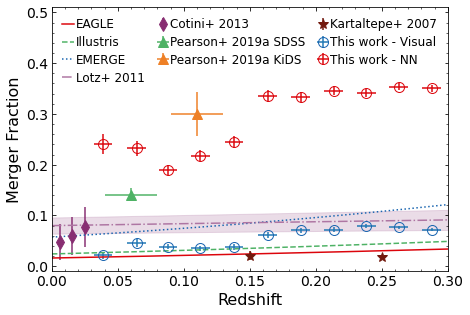}}
	\caption{Merger fraction from the neural networks (red circles) and visual classification (blue circles) as a function of redshift. The merger fraction of the visually confirmed galaxies rises out to $z \approx 0.238$ before falling slightly. The error on the redshift is the standard deviation of the redshift within the redshift bin. The error on the merger fraction is the statistical error. Merger fractions from \citet[][purple diamonds]{2013MNRAS.431.2661C}, \citet[][green and orange triangles]{2019A&A...631A..51P}, \citet[][brown stars]{2007ApJS..172..320K} and \citet[][dot-dashed purple line]{2011ApJ...742..103L} are also shown along with the merger fraction trend from the EAGLE simulation \citep[][solid red line]{2017MNRAS.464.1659Q}, Illustris simulation \citep[][dashed green line]{2015MNRAS.449...49R}, and the pair fraction trend from the EMERGE simulation \citep[][dotted blue line]{2021MNRAS.501.3215O}.}
	\label{fig:merger-frac}
\end{figure}

As a simple application of the catalogue, it is possible to examine how the merger fraction changes as a function of redshift using the visually confirmed mergers. Here we used redshift bins with width 0.025 and determine the mass completeness within each redshift bin as outlined in Sect. \ref{subsec:mass} and shown in Fig. \ref{fig:masslim}. Once the sample of galaxies within each redshift bin is mass complete, we selected redshift bins with more than 100 galaxies and determined the merger fraction for these bins using the visually confirmed mergers. Errors on the merger fractions are Poisson binomial errors. These results can be found in Fig. \ref{fig:merger-frac}. As can be seen, the merger fraction generally rises from 2.1$\pm$0.7\% at $z \approx 0.039$ to 7.9$\pm$0.5\% at $z \approx 0.238$. However, between $z \approx 0.088$ and $z \approx 0.138$, the merger fraction appears to plateau as well as at redshifts above $z \approx 0.238$. Thus generally speaking, mergers are more common in the earlier Universe than we see in the later Universe. This is consistent with theoretical works \citep[e.g.][]{2010ApJ...715..202H, 2010ApJ...724..915H}.

An increasing merger fraction with redshift is consistent with other observational works. Using mergers identified by a CNN, \citet{2019A&A...631A..51P} find an increasing merger fraction as redshift increases, over $0.0 < z < 4.0$ using data from the Sloan Digital Sky Survey \citep{2000AJ....120.1579Y}, KiDS and the Cosmic Assembly Near-infrared Deep Extragalactic Legacy Survey \citep{2011ApJS..197...35G, 2011ApJS..197...36K}. An increase in the merger fraction with redshift is also seen with close pairs, galaxies with projected separations between 5 and 20~kpc, from $z = 0.1$ to $z = 1.2$ \citet{2007ApJS..172..320K}. Using non-parametric statistics, \citet{2013MNRAS.431.2661C} also find that the merger fraction increases with redshift at $z < 0.03$. Similar results were found by \citet{2011ApJ...742..103L}, finding that the fraction of mergers and the fraction of close pair galaxies increases with redshift. \citet{2011ApJ...742..103L} use a Gini-M$_{20}$ cut, asymmetry cut and select close pairs in \textit{Hubble Space Telescope} data for galaxies with stellar masses above $10^{10}$~M$_{\odot}$. 

We converted the merger rates of \citet{2011ApJ...742..103L} into a merger fraction using their merger observability timescale of 0.2~Gyr for comparison with our results. We present their extrapolation to lower redshifts used in this work in Fig. \ref{fig:merger-frac} as the dot-dashed purple line, with their errors shown by the purple shaded region. At higher redshifts, the visually selected mergers are in agreement with the \citet{2011ApJ...742..103L} merger fractions. At lower redshifts, the visually selected merger fraction is lower than that of \citet{2011ApJ...742..103L}. This may be due to the extrapolation required to reach these lower redshifts as the lowest redshift data point of \citet{2011ApJ...742..103L} is at $z = 0.3$. The observability timescale of \citet{2011ApJ...742..103L} has slight redshift dependence which is not presented in the paper. Thus the use of constant timescale may be causing an increase in the \citet{2011ApJ...742..103L} merger fraction presented here at lower redshifts. The merger candidate fraction is much higher than the \citet{2011ApJ...742..103L} merger fraction. As we expect the merger candidates to be contaminated with a large number of non-mergers, this is expected.

\citet{2019A&A...631A..51P} has a much higher merger fraction than this work. This is likely a result of their pure CNN identification of galaxy mergers leaving many false merger detections in the merger sample. This will increase the merger fraction due to the prevalence of non-merging galaxies in the Universe compared to merging galaxies, hence there being more false merger detections than false non-merger detections. Indeed, the merger fraction from the KiDS sample in \citet{2019A&A...631A..51P} is consistent with the merger candidate fraction found by the neural network in this work, before visual confirmation. This consistency between the merger fractions found only with neural networks and these fractions being much larger than the visually selected merger fractions is a strong indication that merger identifications from current neural networks are highly contaminated with non-mergers.

The merger fractions of \citet{2013MNRAS.431.2661C} are larger than the visually inspected merger fractions found in this work. The mergers presented in \citet{2013MNRAS.431.2661C} have been visually checked, like in this work, so there are unlikely to be misclassified non-mergers. However, the size of the merger and non-merger samples are small, a few tens of non-mergers and a few mergers, so these fractions may suffer from low number statistics and so have large uncertainties as seen in Fig. \ref{fig:merger-frac}. \citet{2007ApJS..172..320K} find merger fractions that are lower than this work, as can be seen in Fig. \ref{fig:merger-frac}. As \citet{2007ApJS..172..320K} use close pairs, it is possible that earlier stage mergers are missed that the hybrid neural-network - human classification can find. To add to this, the close pair method misses mergers that are coalescence and post-coalescence which can be detected by the method presented in this work. As a result, it would be expected that the merger fractions presented here are larger than those in \citet{2007ApJS..172..320K}.

Simulations also provide similar results. \citet{2021MNRAS.501.3215O} find the merger fraction increasing with redshift, at least at $z < 1$, in the EMERGE cosmological simulation \citep{2018MNRAS.477.1822M}. This is also seen in the Illustris \citep{2014MNRAS.444.1518V, 2015MNRAS.449...49R} and the Evolution and Assembly of Galaxies and their Environments \citep[EAGLE;][]{2015MNRAS.446..521S, 2017MNRAS.464.1659Q} hydrodynamical, cosmological simulations. However, the Horizon-AGN cosmological simulation \citep{2014MNRAS.444.1453D} finds no evolution of the merger fraction with redshift \citep{2015MNRAS.452.2845K}.

The results presented in this paper show similar merger fractions to the EAGLE results for galaxies with stellar masses above 10$^{10}$~M$_{\odot}$, at the lower redshifts presented in this work. At higher redshifts we find higher merger fractions than EAGLE. Higher merger fractions than EAGLE are expected. As our mass limits are lower than EAGLE at all redshifts, the largest mass limit in this work is less than 10$^{9}$~M$_{\odot}$, we expect to find higher merger fractions \citep{2013MNRAS.430.1158S}. This is likely due to a greater fraction of lower mass galaxies are undergoing a merger compared to higher mass galaxies \citep{2013MNRAS.430.1158S, 2014MNRAS.445.1157C, 2020A&A...644A..87W}. If the redshift evolution of the merger fraction in EAGLE and the visually selected mergers in this work were the same, it would be expected that the merger fraction of EAGLE and this work would converge towards higher masses, where the mass limit of this work becomes closer to that of EAGLE. As the opposite is seen, we find a faster increase in merger fraction with redshift compared to EAGLE if the same mass limits are used.

The Illustris simulation shows a similar merger fraction trend to the EAGLE simulation. Using the merger rate given by \citet{2015MNRAS.449...49R} and assuming a merger ratio of 1/4\footnote{larger merger ratios increase the merger fraction} and a descendant mass of 10$^{10}$~M$_{\odot}$\footnote{larger descendent masses increase the merger fraction}, the merger rate is converted to a merger fraction by assuming an average observational timescale (T$_{obs}$) of 0.65~Gyr \citep{2011ApJ...742..103L}. The Illustris merger fraction is consistent with the visually selected merger fraction at $z < 0.15$. The visually selected merger fraction increases at a faster rate with redshift than the Illustris merger fraction and so the visually selected merger fraction rises above the Illustris merger fraction at $z > 0.15$.

The EMERGE simulation finds a larger pair fraction than the visually selected merger fraction. In Fig. \ref{fig:merger-frac}, we show the pair fraction from \citet{2021MNRAS.501.3215O} for simulated galaxy pairs with M$_{\star} \geq 10^{10.3}$~M$_{\sun}$ and projected distances between 5 and 50~kpc. As with the Illustris merger rate, the EMERGE merger rate has been multiplied by a T$_{obs}$ of 0.65~Gyr to get the merger fraction. The EMERGE merger fraction is slightly larger than the visually selected merger fraction of this work, as can be seen in Fig. \ref{fig:merger-frac}. However, the merger rate derived in \citet{2021MNRAS.501.3215O} does not apply a correction factor to account for not all galaxy pairs resulting in a merger. If a typical correction factor of 0.6 is applied \citep[e.g.][]{2014ARA&A..52..291C}, the EMERGE merger fraction is in good agreement with the visually selected merger fraction of this work. On the other hand, as pair samples miss post-coalescence galaxies, approximately half of all mergers, the EMERGE pair fraction as presented in \citet{2021MNRAS.501.3215O} is likely to be close to the true merger fraction of pre and post-merger galaxies. This is a result of the factor of 0.6 reduction not applied in \citet{2021MNRAS.501.3215O} being almost entirely offset by the approximate factor of 2 needed to account for the missing post-coalescence galaxies.

The comparisons with other works presented here are not exhaustive. There is a wealth of similar studies onto the merger fraction and merger rates in the Universe from a number of different surveys and data sources. They do, however, typically all indicate that the merger fraction increases with redshift, although the evolution with redshift does differ \citep[e.g.][]{2002ApJ...565..208P, 2004ApJ...617L...9L, 2009A&A...498..379D, 2013A&A...553A..78L, 2014MNRAS.445.1157C, 2017MNRAS.470.3507M, 2020ApJ...895..115F}. Moreover, \citet{2008MNRAS.386..909C} has an increasing merger fraction at lower redshifts but a decreasing fraction at high redshifts of $z \gtrapprox 2.5$.

\subsection{Star-formation Enhancement}
A second simple application of the catalogue allows the study of the star-formation enhancement, or lack thereof, due to galaxy mergers. For this, we compared the SFRs in the merger candidates and visually verified mergers with the SFRs in a control sample for star-forming galaxies.

\begin{figure}
	\resizebox{\hsize}{!}{\includegraphics{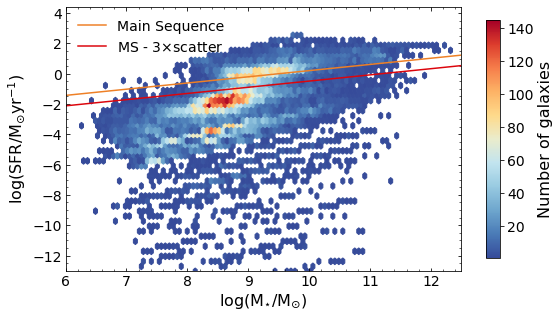}}
	\caption{M$_{\star}$-SFR plane for the galaxies classified by the neural networks. The number of galaxies in each M$_{\star}$-SFR bin is from low in blue to high in red. The orange line indicates the \citet{2018A&A...615A.146P} main-sequence at $z = 0.24$ while the red line indicated three times the scatter below the main-sequence. Galaxies are classified as star-forming if they lie above the red line.}
	\label{fig:mainsequence}
\end{figure}

Star-forming galaxies were selected based on their position relative to the galaxy main-sequence (MS), a tight correlation between the M$_{\star}$ and SFR of star-forming galaxies \citep[e.g.][]{2004MNRAS.351.1151B, 2007ApJ...660L..43N, 2007A&A...468...33E, 2014ApJS..214...15S}. We use the redshift dependent main-sequence of \citet{2018A&A...615A.146P} at $z = 0.24$ (orange line in Fig.\ref{fig:mainsequence}), the mean redshift of the mass complete sample defined in Sect. \ref{subsec:mgr-frac}. The empirical scatter in this MS is $\sigma_{MS} = 0.23$~dex. We consider galaxies above MS$ - 3\sigma_{MS}$ as part of the MS (red line in Fig. \ref{fig:mainsequence}).

\begin{figure*}
	\resizebox{0.48\hsize}{!}{\includegraphics{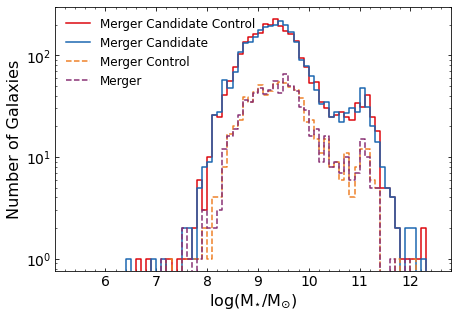}}
	\resizebox{0.5\hsize}{!}{\includegraphics{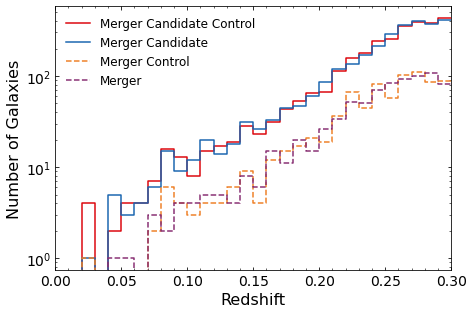}}
	\caption{Stellar mass (left panel) and redshift (right panel) distributions for merger candidates (blue) and visually confirmed mergers (dashed purple) and their selected non-merging control galaxies (red and dashed orange for candidate and visually confirmed merger controls, respectively).}
	\label{fig:control}
\end{figure*}

The control galaxies were selected from the non-merging samples, defined as galaxies with $\texttt{frac\_merger} < 0.5$ for both the merger candidates and the visually selected mergers. For each merger (candidate), non-merging galaxies within 0.05~dex in M$_{\star}$ and 0.005 in redshift were identified. Where at least one matching non-merger was found, each merger (candidate) was assigned a unique non-merger control. If no match was found, the matching distance was then increased by 0.05~dex in M$_{\star}$ and 0.005 in redshift and the matching process repeated for any unmatched merging galaxies. The matching distance was repeatedly increased until the M$_{\star}$ matching distance was over 0.3~dex or the redshift matching distance was over 0.05, the typical weighted dispersion of the photometric redshifts. Any merger (candidates) that had not been matched were then removed. Matching was done independently for the merger candidates and visually confirmed mergers. This process resulted in 2\,905 of the 3\,342 mass complete merger candidates and 801 of the 803 mass complete mergers having a matched control galaxy. The mass and redshift distributions for the merger candidates, mergers and their respective controls are shown in Fig. \ref{fig:control}. The environment in which a galaxy lies, for example in a group, cluster or the field, can influence the SFR of a galaxy independent of if the galaxy is interacting or not, with high density regions having lower typical SFR \citep[e.g.][]{2002MNRAS.334..673L, 2010ApJ...721..193P, 2020MNRAS.499..631V}. For this work, the envirnoment in which the galaxy mergers and non-merger controlls lie was not considered when matching mergers with thier controlls.

We find that for the merger candidates, the SFR of the control sample is only slightly lower than that of the merger candidate sample. Here we subtract the log(SFR/M$_{\odot}$yr$^{-1}$) of the control galaxies from the log(SFR/M$_{\odot}$yr$^{-1}$) of their matched merger candidate. The average of these differences is
\begin{equation}
	\Delta\log(\mathrm{SFR}/M_{\odot}yr^{-1}) = 0.071 \pm 0.014~\mathrm{dex}
\end{equation}
with a sample standard deviation in $\Delta\log(\mathrm{SFR}/M_{\odot}yr^{-1})$ of
\begin{equation}
	\sigma\Delta\log(\mathrm{SFR}/M_{\odot}yr^{-1}) = 0.733 \pm 0.010~\mathrm{dex}.
\end{equation}
The errors on $\sigma\Delta\log(\mathrm{SFR}/M_{\odot}yr^{-1})$ are the standard error for the standard deviation. Thus, the average SFR enhancement is in 5$\sigma$ tension of being zero. The average enhancement of the visually confirmed merger sample, derived in the same manner, is smaller than that of the merger candidates, with an average change to log(SFR/M$_{\odot}$yr$^{-1}$) being 
\begin{equation}
	\Delta\log(\mathrm{SFR}/M_{\odot}yr^{-1}) = 0.040 \pm 0.025~\mathrm{dex}
\end{equation}
and a sample standard deviation of
\begin{equation}
	\sigma\Delta\log(\mathrm{SFR}/M_{\odot}yr^{-1}) = 0.696 \pm 0.017~\mathrm{dex}.
\end{equation}
For the visually selected mergers, therefore, the average enhancement is in 1$\sigma$ tension of being zero.

We know that the merger candidate sample is contaminated with non-mergers and the visually selected sample is also likely to be contaminated. If non-mergers have a lower SFR than merging galaxies, this contamination will act to reduce the SFR enhancement seen when comparing the mergers with non-mergers. The merger candidate and visually confirmed samples are likely to have a large fraction of widely separated galaxies. This is a result of widely separated galaxies being easier to identify by eye, and the training sample being based on visual selection and the visual confirmation being visual by design. Closely separated galaxies may show more merger-like features, but two very close galactic cores can be indistinguishable from a single galactic core. This makes the choice between a merging galaxy or an irregular galaxy difficult. More closely separated galaxies are more likely to have higher SFRs \citep[e.g.][]{2015MNRAS.452..616D, 2019MNRAS.485.1320M} so a high fraction of widely spaced galaxies can weaken any SFR enhancement. To add to this, if our control sample is primarily selected from lower density environments while the mergers are in higher density environments, this will act to suppress the apparent SFR enhancement for the merging systems. These factors could be combining to give such a marginal enhancement in both samples.

These results are qualitatively in line with other works in that the enhancement we see to SFR is less than a factor of two. The low SFR enhancement is consistent with \citet{2018ApJ...868...46S}, who find no significant SFR enhancement in merging galaxies when compared to non-mergers. \citet{2015MNRAS.454.1742K} find a typical increase in SFR by up to a factor of 1.9 for the most highly interacting and closest pair galaxies with a reduction in SFR enhancement as the galaxies are more wideley separated. \citet{2011A&A...535A..60H} also see an increase in SFR of a factor of approximately 2 for merging galaxies when compared to non-merging counterparts. \citet{2013MNRAS.435.3627E} find an increase in SFR by a factor of 2 for pre-merger galaxies and 3.5 for post mergers. They also find that the enhancement is greater for galaxy pairs with smaller separation. In this work, we do not distinguish between pre and post-mergers nor do we determine the separation of the galaxy pairs. However, as the enhancement of the merger galaxies is much less than the 3.5 seen for post-merger galaxies in \citet{2013MNRAS.435.3627E}, it suggests that the sample of galaxies herein identified are primarily pre-merger galaxies. The visual confirmation of the galaxy mergers did not directly record the type of merger, pre-merger or post-merger, but a brief reinspection shows the identified galaxies are more likely to be pre-merger systems. The visual identifier WJP is not biased towards pre-merger or post-merger systems while LES has a slight bias towards post-merger systems, see Appendix \ref{app:authors}, suggesting the greater number of pre-mergers is not a result of visual classifier selection bias. The larger SFR enhancement seen in \citet{2013MNRAS.435.3627E} may be a result of their controlling for environment. The environment controlled sample of \citet{2013MNRAS.435.3627E} will not suffer from the potential environmental influence that may be influencing our results, as noted above.

The work of \citet{2019A&A...631A..51P} provides an interesting comparison to this work. Both this work and that of \citet{2019A&A...631A..51P} identify galaxy mergers using neural networks, a CNN in \citet{2019A&A...631A..51P} and a CNN-FCN-human hybrid here. \citet{2019A&A...631A..51P} find a typical enhancement to SFR of a factor of $1.15 \pm 0.12$ while we find an enhancement of a factor of $1.178 \pm 0.065$ for the merger candidates, the closest comparison. Evidently, these two values are consistent within the standard errors in the means. Both the \citet{2019A&A...631A..51P} merger sample and the merger candidates of this work are likely to be contaminated with non-merging galaxies. If mergers are typically of higher SFR than non-mergers it would be expected that the increase in SFR for a merger sample with fewer contaminants would have a larger change in SFR. However, this is not what is seen in this work, indeed a smaller enhancement to SFR is seen with the visually confirmed merger sample. Thus, it may be expected that cleaning the \citet{2019A&A...631A..51P} merger sample may similarly see a smaller SFR enhancement. We note, however, that \citet{2019A&A...631A..51P} compare the average SFR of their mergers with the average SFR of the non-mergers while here we compare the difference in SFR between a merger and its matched control.

\section{Summary}\label{sec:conc}
In this paper, we present a catalogue of galaxy mergers in the North Ecliptic Pole field using optical data from the Hyper Suprime-Cam. The merger identification is a hybrid of automated and human classification: a neural network is used to identify merger candidates which are then visually inspected. The neural networks used a combination of both images and morphological parameters, which was found to provide better results than just images or morphological parameters alone. From the hybrid approach for merger identification, the final catalogue contains 2\,109 merging galaxies out of a total of 34\,264 galaxies with redshifts between 0.0 and 0.3.

Studying the networks and how they classify the galaxies, it appears that both the networks will miss-classify galaxies as merging that have a companion that is close in projection but not physically associated. Both networks appear to primarily rely on the morphological parameters for classification with parameters that examine the asymmetry of the light distribution being a key component.

As test applications of the catalogue, we performed an analysis of the merger fraction as a function of redshift and examine the SFR enhancement due to galaxy mergers. We find that the evolution of the merger fraction is qualitatively consistent with merger fraction evolutions found in other observational surveys as well as cosmological simulations. For the SFR enhancement, we find a mild increase by a factor of $1.178 \pm 0.065$ for the merger candidates and a factor of $1.096 \pm 0.063$ for the visually confirmed mergers, consistent with other works.

The resulting catalogue is well placed to be exploited for further use within the NEP field. It is also in a prime position to be used as a training set for the upcoming \textit{Euclid} Northern deep field. Due to the scale of the upcoming surveys, such a catalogue could prove to be invaluable as a training sample for automated merger detection over larger regions of the sky.

\begin{acknowledgements}
We would like to thank the referee for their thorough and thoughtful comments that helped improve the quality and clarity of this paper.
	
We would like to thank C.~Conselice, A.~Graham, A.~Nanni and D.~J.~D.~Santos for helpful discussions on this paper.
	
W.J.P. has been supported by the Polish National Science Center project UMO-2020/37/B/ST9/00466.

K.M has been supported by the Polish National Science Center project UMO-2018/30/E/ST9/00082.

This research was conducted under the agreement on scientific cooperation between the Polish Academy of Sciences and the Ministry of Science and Technology in Taipei and supported by the Polish National Science Centre grant UMO-2018/30/M/ST9/00757 and by Polish Ministry of Science and Higher Education grant DIR/WK/2018/12.

The Hyper Suprime-Cam (HSC) collaboration includes the astronomical communities of Japan and Taiwan, and Princeton University. The HSC instrumentation and software were developed by the National Astronomical Observatory of Japan (NAOJ), the Kavli Institute for the Physics and Mathematics of the Universe (Kavli IPMU), the University of Tokyo, the High Energy Accelerator Research Organization (KEK), the Academia Sinica Institute for Astronomy and Astrophysics in Taiwan (ASIAA), and Princeton University. Funding was contributed by the FIRST programme from the Japanese Cabinet Office, the Ministry of Education, Culture, Sports, Science and Technology (MEXT), the Japan Society for the Promotion of Science (JSPS), Japan Science and Technology Agency (JST), the Toray Science Foundation, NAOJ, Kavli IPMU, KEK, ASIAA, and Princeton University. 

This paper makes use of software developed for the Large Synoptic Survey Telescope. We thank the LSST Project for making their code available as free software at  http://dm.lsst.org

This paper is based [in part] on data collected at the Subaru Telescope and retrieved from the HSC data archive system, which is operated by Subaru Telescope and Astronomy Data Center (ADC) at National Astronomical Observatory of Japan. Data analysis was in part carried out with the cooperation of Center for Computational Astrophysics (CfCA), National Astronomical Observatory of Japan.

GAMA is a joint European-Australasian project based around a spectroscopic campaign using the Anglo-Australian Telescope. The GAMA input catalogue is based on data taken from the Sloan Digital Sky Survey and the UKIRT Infrared Deep Sky Survey. Complementary imaging of the GAMA regions is being obtained by a number of independent survey programmes including GALEX MIS, VST KiDS, VISTA VIKING, WISE, Herschel-ATLAS, GMRT and ASKAP providing UV to radio coverage. GAMA is funded by the STFC (UK), the ARC (Australia), the AAO, and the participating institutions. The GAMA website is http://www.gama-survey.org/ .

Based on observations made with ESO Telescopes at the La Silla Paranal Observatory under programme ID 177.A-3016.
\end{acknowledgements}

\bibliographystyle{aa} 
\bibliography{425-merger-catalogue-NEP} 

\begin{appendix}
\section{CNN performance definitions}\label{app:definitions}
The terms used to describe the performance of the neural networks presented in this work may be an alternate nomenclature to other works or may be unfamiliar. To avoid confusion we present the definitions used in this work in Table \ref{table:definitions}.

\begin{table*}
	\caption{Terms used when describing the performance of neural networks from \citet{2019A&A...626A..49P}}
	\begin{center}
		\begin{tabular}{p{0.16\textwidth}p{0.54\textwidth}p{0.23\textwidth}}
			\hline
			Term & Definition & \\
			\hline
			True Positive (TP) & An object known to be a merger that is identified by a network as a merger. & \\
			False Positive (FP) & An object known to be a non-merger that is identified by a network as a merger. & \\
			True Negative (TN) & An object known to be a non-merger that is identified by a network as a non-merger. & \\
			False Negative (FN) & An object known to be a merger that is identified by a network as a non-merger. & \\
			Recall & Fraction of objects correctly identified by a network as a merger with respect to the total number of objects classified in the catalogues as mergers. & TP / (TP+FN) \\ 
			Specificity & Fraction of objects correctly identified by a network as a non-merger with respect to the total number of objects classified in the catalogues as non-mergers. & TN / (TN+FP) \\
			Precision & Fraction of objects correctly identified by a network as a merger with respect to the total number of objects identified by a network as a merger. & TP / (TP+FP) \\
			Negative Predictive Value (NPV) & Fraction of objects correctly identified by a network as a non-merger with respect to the total number of objects identified by a network as a non-merger. & TN / (TN+FN) \\
			Accuracy & Fraction of objects, both merger and non-merger, correctly identified by a network. & (TP+TN) / (TP+FP+TN+FN) \\
			\hline
		\end{tabular}
		\label{table:definitions}
	\end{center}
\end{table*}

\section{Visual inspection performance}\label{app:authors}
To check the performance of the two authors' merger identification, images of mergers and non-mergers from the Illustris TNG simulation \citep{2018MNRAS.480.5113M, 2018MNRAS.477.1206N, 2018MNRAS.475..624N, 2018MNRAS.475..648P, 2018MNRAS.475..676S, 2019ComAC...6....2N} were used. A sample of 100 major merger galaxies were selected, along with a further 100 non-mergers, from snapshot 87 ($z = 0.15$). Here, a major merger is defined to have merged in the last 500~Myr or will merge in the next 1000~Myr and have a mass ratio of $< 4:1$ \citep{2020A&A...644A..87W}. This mass ratio is derived from the stellar masses of the two merging galaxies at the snapshot when the secondary galaxy reached its maximum stellar mass \citep{2015MNRAS.449...49R}. The simulated galaxies were then convolved with the point-spread function of the HSC-NEP images before being embedded into the HSC-NEP images to add realistic noise and chance projections. The position in the image where the simulated galaxies were embedded were selected such that there were no sources in the HSC-NEP catalogue within 10~arcsec (64~pixels). These mock galaxy observations were then classified by WJP and LES, with neither knowing if the galaxy was truly a merger or non-merger, only that the sample had an equal number of each. The results of the performance test can be found in Table \ref{tab:authors}.

\begin{table}
	\caption[]{Performance statistics of WJP and LES classifying simulated observations of merging and non-merging galaxies.}
	\label{tab:authors}
	\centering
	\begin{tabular}{lcc}
		\hline
		Statistic & WJP & LES\\
		\hline
		Accuracy & 0.620 & 0.630\\
		Recall & 0.450 & 0.440\\
		Precision & 0.682 & 0.710\\
		Specificity & 0.790 & 0.820\\
		NPV\tablefootmark{a} & 0.590 & 0.594\\
		\hline
	\end{tabular}
	\tablefoot{
		\tablefoottext{a}{Negative predictive value}\\
		Defenitions of the statistics can be found in Appendix \ref{app:definitions}.
	}
\end{table}

\begin{figure}
	\resizebox{\hsize}{!}{\includegraphics{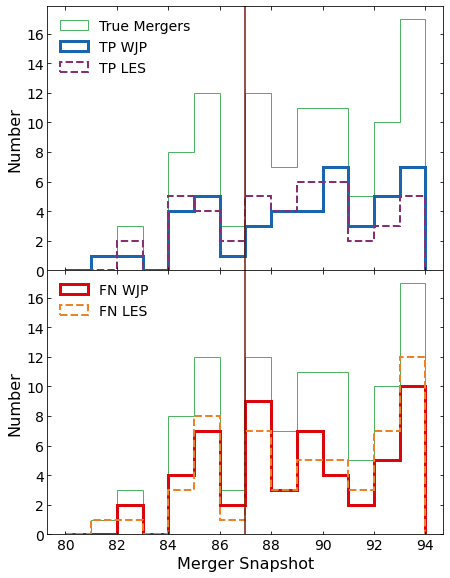}}
	\caption{Plot of correctly and incorrectly identified images of simulated mergers as a function of snapshot when the merger occurred. The blue and red lines indicate the TP and FN, respectively, classified by WJP while the purple and orange dashed lines indicate the TP and FN, respectively, classified by LES. The Green line indicates the total number of mergers used from Illustris TNG while the vertical brown line indicates the snapshot of observation (87, $z = 0.15$).}
	\label{fig:authors}
\end{figure}

As can be seen, the performance of WJP and LES is lower than that of the neural networks presented in this work but both are similar. WJP has fewer FP than LES (21 compared to 18) but also has more TP (45 compared to 41). However, both authors correctly identify fewer than half of the merging galaxies. For the FN, there is no clear trend with time before or after the merger, as shown in Fig. \ref{fig:authors}. WJP has a larger fraction of missed mergers that are close to the merger event (here defined as the snapshot when two galaxies are tracked as one in the simulation) than LES while LES has a larger fraction of missed mergers with longer times until the merger event will take place. Thus WJP is likely to miss merging galaxies that are physically close to one another or have just merged. On the other hand, LES is likely to miss galaxies that are at the early stages of a merger.

\section{Examples of differet galaxy classification}\label{app:random}
\begin{figure}
	\centering
	\resizebox{\hsize}{!}{\includegraphics{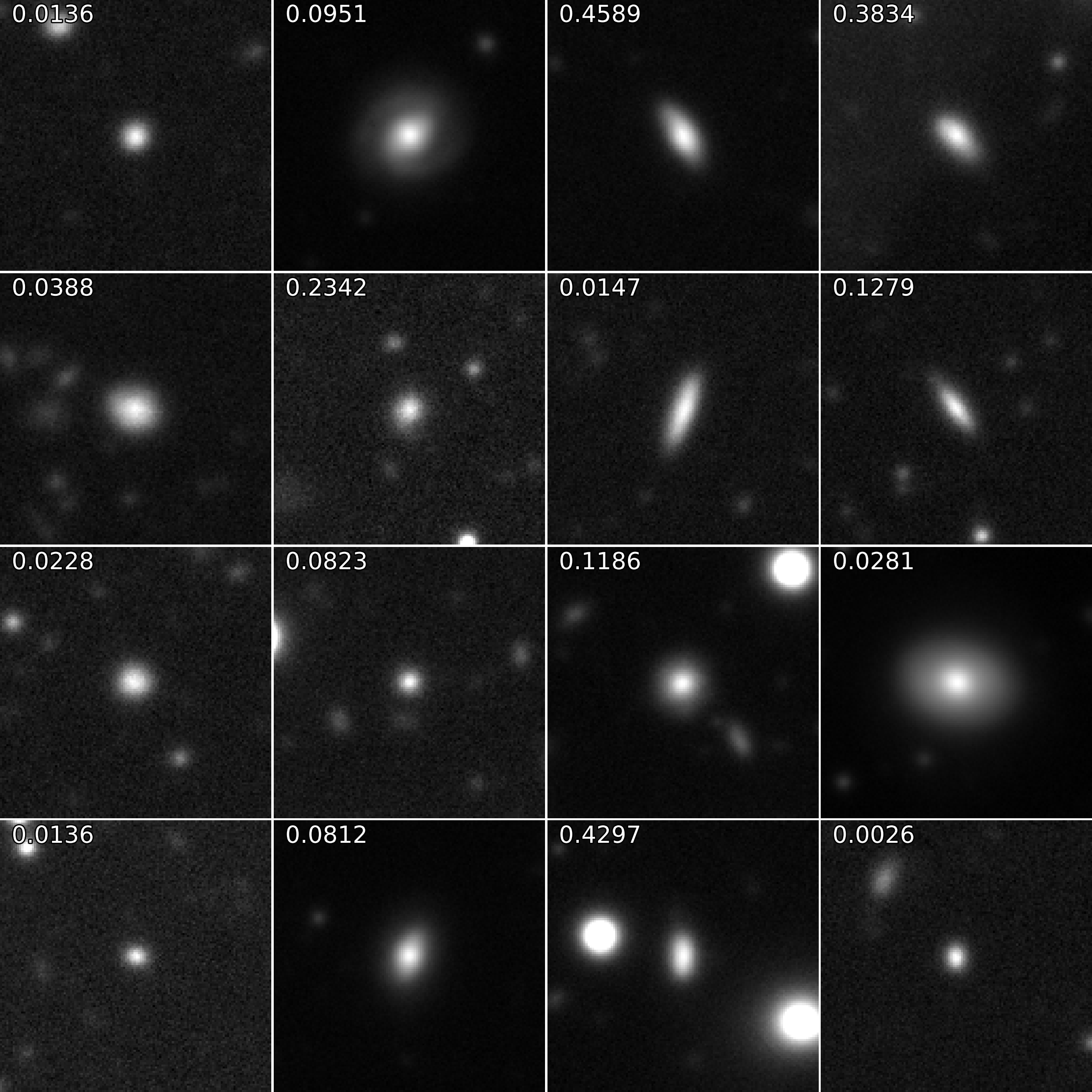}}
	\caption{Sixteen randomly selected galaxies identified as non-mergers by the neural networks with \texttt{frac\_merger} in the upper left corner of the image.}
	\label{fig:random-nonmerger}
\end{figure}

\begin{figure}
	\centering
	\resizebox{\hsize}{!}{\includegraphics{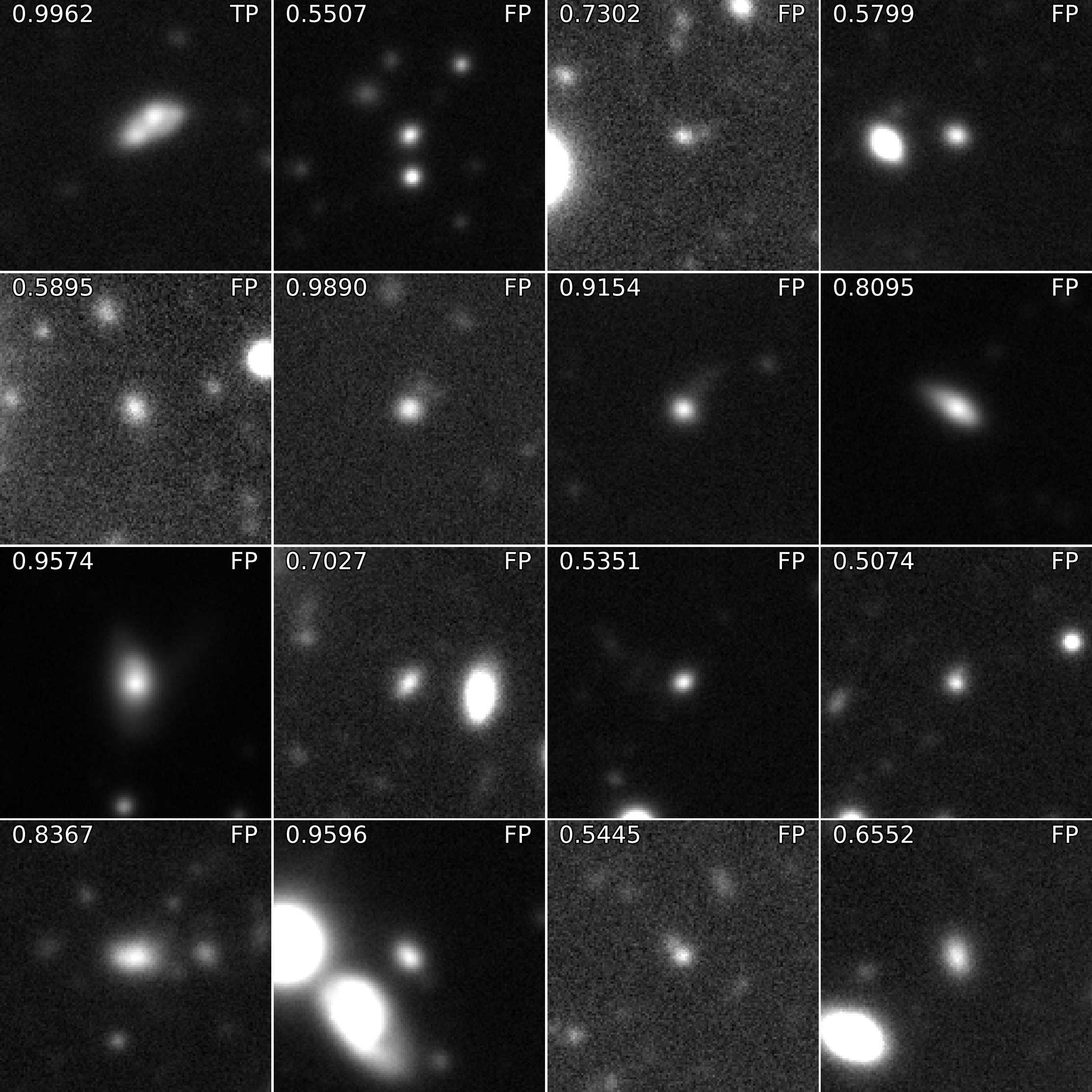}}
	\caption{Sixteen randomly selected galaxies identified as mergers by the neural networks with \texttt{frac\_merger} in the upper left corner of the image and if they are TP or FP in the top right corner.}
	\label{fig:random-merger}
\end{figure}

\begin{figure}
	\centering
	\resizebox{\hsize}{!}{\includegraphics{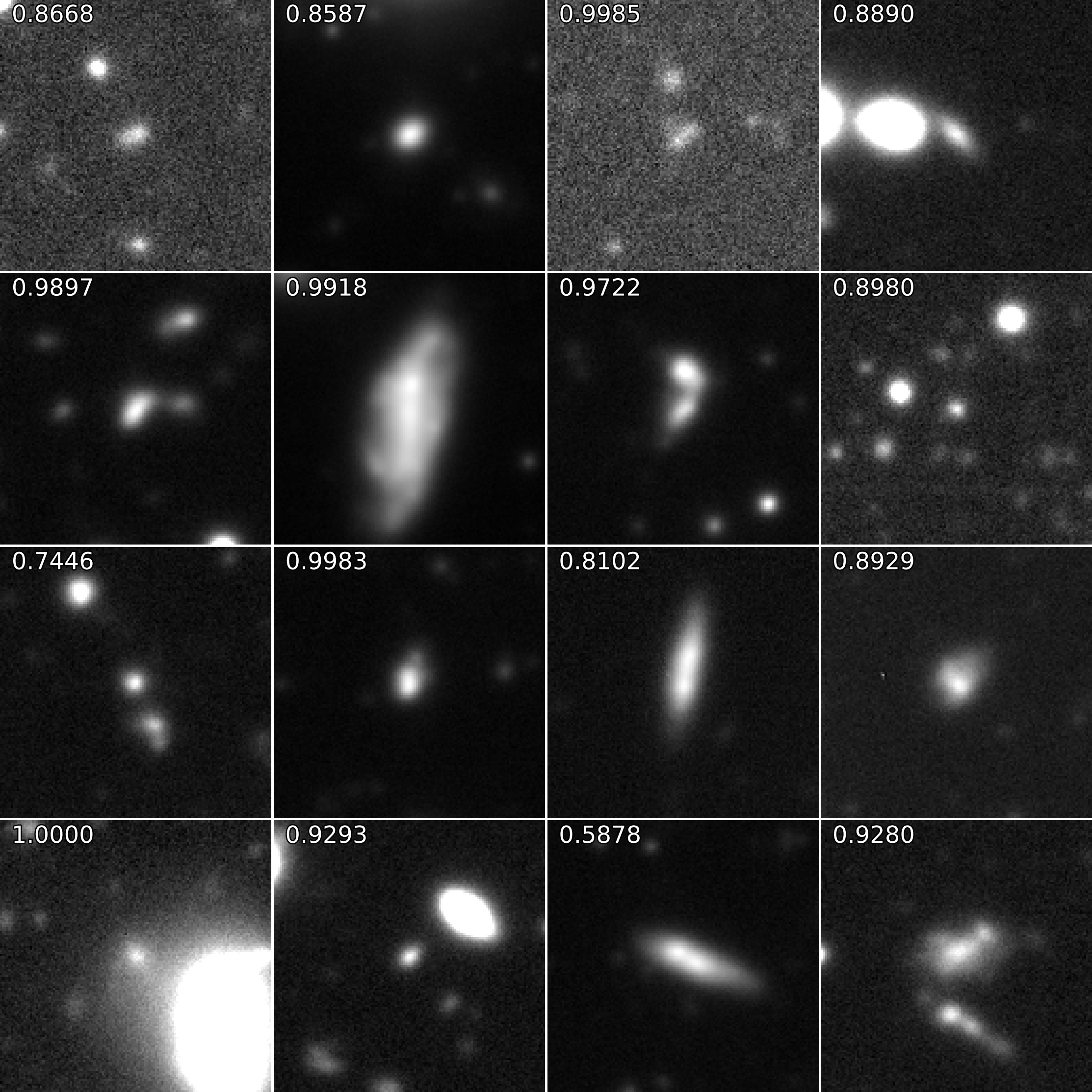}}
	\caption{Sixteen randomly selected galaxies visually confirmed to be mergers with \texttt{frac\_merger} in the upper left corner of the image.}
	\label{fig:random-visual}
\end{figure}

Further to the discussion in the main body of the paper, in this appendix we study, in more detail, if the neural network has been inadvertantly trained to identify structured and unstructured galaxies. To this end, we present 16 randomly selected non-mergers in Fig. \ref{fig:random-nonmerger} and 16 randomly selected merger candidates (both TP and FP) in Fig. \ref{fig:random-merger}. While the non-mergers presented in Fig. \ref{fig:random-nonmerger} are predominantly unstructured galaxies, the same is also true of the merger candidates presented in Fig. \ref{fig:random-merger}. Examples of structured non-mergers are presented in Fig. \ref{fig:structure} above. The images of unstructured merger candidates do typically contain other galaxies or stars in close projection, although later may be found to be unassociated during visual inspection. If the neural networks had been trained to identify structured and unstructured galaxies, very few merger candidates would be expected to not have structure regardless of whether there are other objects in close projection. As this is not the case, we conclude that the networks are indeed identifying mergers and non-mergers as intended. For comparison, we also present 16 randomly selected visually confirmed mergers in Fig. \ref{fig:random-visual}.

\end{appendix}

\end{document}